\documentstyle[12pt,epsfig,aps]{revtex}

\newcommand{\dfrac}{\displaystyle\frac}
\begin{document}

\draft

\title{Study of Proton Magic Even-Even Isotopes and Giant Halos of Ca Isotopes with
Relativistic Continuum Hartree-Bogoliubov Theory}
\author{S.Q. Zhang$^{a}$, J. Meng$^{a,b,c}$\thanks{e-mail: mengj@pku.edu.cn}, H. Toki$^{d}$, I. Tanihata$^{e}$,
S.-G. Zhou$^{a,b,c}$}
\address{${}^{a}$School of Physics, Peking University, Beijing 100871, P.R. China}
\address{${}^{b}$Institute of Theoretical Physics, Chinese Academy of Sciences, Beijing 100080, P.R. China}
\address{${}^{c}$Center of Theoretical Nuclear Physics, National Laboratory of
       Heavy Ion Accelerator, Lanzhou 730000, P.R. China}
\address{${}^{d}$Research Center for Nuclear Physics, Osaka University (RCNP), Ibaraki, Osaka 567-0047, Japan}
\address{${}^{e}$The Institute of Physical and Chemical Research (RIKEN),
       Hirosawa 2-1, Wako, Saitama 351-0198, Japan}
\date{\today}
\maketitle

\begin{abstract}
We study the proton magic O, Ca, Ni, Zr, Sn, and Pb isotope chains
from the proton drip line to the neutron drip line with the
relativistic continuum Hartree-Bogoliubov (RCHB) theory.
Particulary, we study in detail the properties of even-even Ca
isotopes due to the appearance of giant halos in neutron rich Ca
nuclei near the neutron drip line. The RCHB theory is able to
reproduce the experimental binding energies $E_b$ and two neutron
separation energies $S_{2n}$ very well. The predicted neutron drip
line nuclei are $^{28}$O, $^{72}$Ca, $^{98}$Ni, $^{136}$Zr,
$^{176}$Sn, and $^{266}$Pb, respectively. Halo and giant halo
properties predicted in Ca isotopes with $A>60$ are investigated
in detail from the analysis of two neutron separation energies,
nucleon density distributions, single particle energy levels, the
occupation probabilities of energy levels including continuum
states. The spin-orbit splitting and the diffuseness of nuclear
potential in these Ca isotopes are studied also. Furthermore, we
study the neighboring lighter isotopes in the drip line Ca region
and find some possibility of giant halo nuclei in the Ne-Na-Mg
drip line nuclei.
\end{abstract}


\section{Introduction}
The study of exotic nuclei, which are so-called due to their large
N/Z ratios (isospin) and their interesting properties, e.g. halo
and skin, has attracted world wide attention \cite{TA95,MU01}.
With the recent developments in accelerator technology and
detection techniques, it has come into reality to produce these
exotic nuclei and study their detailed properties with the
radioactive ion beam (RIB) facilities. Since Tanihata et al.
discovered the first case of halo in an exotic nucleus $^{11}$Li
with RIB in 1985 \cite{TA85}, more and more exotic nuclei have
been investigated with various modern experimental methods to
understand this attractive phenomenon better. For nuclei far from
the $\beta$-stability valley and with small nucleon separation
energy, the valence nucleons in exotic nuclei extend over quite a
wide space to form low density nuclear matter. It is expected that
the "halo" in exotic nuclei provides an interesting case to study
the nuclear environment in astrophysics in the laboratory.

Very neutron-rich nuclei and, in particular, those near the
neutron-drip lines and near closed shells, play an important role
in nuclear astrophysics. Their properties such as binding
energies, neutron separation energies, deformation parameters,
etc., strongly affect the way neutron-rich stable isotopes are
formed in nature by the so-called {\it r} process.

Furthermore, the exotic nuclei are expected to exhibit some other
interesting phenomena such as the disappearance of traditional
shell gaps and the occurrence of new shell gaps, which result in
new magic numbers. Ref.\cite{OZ00} has reported that the new magic
number $N=16$ appears in the light neutron drip line region. Ozawa
et al. have suggested the mechanism for the formation of the new
magic number, which is intimately related to the neutron halo
formation and is common in nuclei near neutron drip line.

It is very helpful to use self-consistent microscopic model to
study the properties of exotic nuclei. During the past two
decades, the relativistic mean field (RMF) theory has received
wide attention, because the RMF theory is very successful in
describing many nuclear phenomena for nuclei even far from
stability as well as the stable nuclei. Compared with the
non-relativistic mean field theory, RMF can reproduce the right
nuclear incompressibility coefficient and saturation properties
(Coester line) in nuclear matter and gives naturally the
spin-orbit coupling potential. The reviews on RMF theory are given
in Ref. \cite{WA74,SW86,RI96} . The starting point of the RMF
theory is the Lagrangian which describes the nucleons as Dirac
spinors moving in a mean field, composed of the interaction
between nucleons (protons and neutrons) and mesons ($\sigma,
\omega, \rho$), and also Coulomb field. From this viewpoint, the
RMF theory is more microscopic in a sense of describing the
nuclear system at the meson level.

Usually for exotic nuclei, their Fermi surfaces are very close to
the continuum threshold. In these cases, the valence nucleons
could be easily scattered to the continuum states due to the
pairing correlation. Thus, theories which can properly handle the
pairing and continuum states are needed to describe the properties
of exotic nuclei. For the pairing interaction, the simple BCS
method and general Bogoliubov quasi-particle transformation are
two candidates. The former is very useful and successful for
stable nuclei. However, when it is extended to exotic nuclei, the
occupation probability would become finite for those continuum
states and involve unphysical fermion gas  In this case, we should
use the Bogoliubov transformation to handle the pairing
correlation in exotic nuclei instead of the simple BCS method.
Based on the RMF theory, the relativistic Hatree-Bogoliubov (RHB)
equation can be derived by quantization of the meson fields as
well as the nucleon fields in Lagrangian density \cite{KR91}.
Furthermore, in order to describe self-consistently both the
continuum and bound states and the coupling between them, the RHB
theory must be solved in coordinate space, i.e., the newly
developed Relativistic Continuum Hartree-Bogoliubov (RCHB) theory
\cite{MR96,ME98}.

The RCHB theory was extremely successful in describing the ground
state properties of nuclei both near and far from the
$\beta$-stability line. A remarkable success of the RCHB theory is
the new interpretation of the halo in $^{11}$Li \cite{MR96} and
the prediction of the exotic phenomenon as giant halos in Zr
($A>122$) isotopes \cite{MR98}. The giant halos are very
interesting phenomena in exotic nuclei because the halos are
formed by more than two neutrons scattered as Cooper pairs to the
continuum region. However, the exotic Zr ($Z=40$) isotopes with
$N>82$ are rather heavy for observation by the RIB facilities at
the present. With the present RIB techniques, light drip line
nuclei are more probable to be accessed with the transfer reaction
or other reaction mechanism. It is thus extremely valuable for us
to investigate the giant halo phenomena in lighter nuclei like Ca
($Z=20$) isotopes\cite{MT02}.

The Calcium isotope chains have received much attention due to its
rich experimental results on binding energy, density distribution,
single particle energy, radius, etc. Though those data are now
limited near the stability line, it is useful to calculate these
quantities in order to test the microscopic theory by future
experiments. Also, there lie shell effects in this chain due to
the short shell period as the traditional magic numbers are $N = $
20, 28, 50 and sub-magic number $N=40$. The properties of Ca
isotope have been investigated with various methods based on the
mean field theory, e.g., the non-relativistic Hartree-Fock (HF)
and Hartree- Fock Bogoliubov (HFB) method \cite{FTZ00}, the Skyrme
Hartree Fock (SHF) method \cite{IM00}, relativistic density
dependent Hartree-Fock method \cite{CH99}, and the self-consistent
Hartree-Fock calculation plus the random-phase approximation
\cite{HSZ01}. We apply here the RCHB theory to investigate the
ground state properties of the Ca isotope chain, especially
probing the halo properties in the exotic nuclei near the neutron
drip line.

In the former letter \cite{MT02}, we have reported briefly the
halos discovered in Ca isotopes near the drip line region with the
RCHB theory. Here we give details of the investigation of the
mechanism of the appearance of the giant halos. Besides the
detailed discussion on Ca isotopes, the ground state properties of
other proton magic isotope chains, O, Ni, Zr, Sn, and Pb are
discussed in this article. The paper is organized as follows: In
Sec. II, a brief outline of the RCHB theory is given. In Sec. III,
we provide the numerical results on these proton magic even-even
nuclei and the miscellaneous properties for Ca isotopes, such as
two neutron separation energies $S_{2n}$, radius, density
distribution, single particle energy level, contribution from
continua, spin-orbit splitting, potential diffuseness. In Sec. IV
the prospect on the theoretical progress and experimental
expectation of the giant halo nuclei are reviewed. Finally, Sec. V
summarizes our main results.

\section{Relativistic Continuum Hartree Bogoliubov Theory}

The RCHB theory, which is the extension of the RMF theory with the
Bogoliubov transformation in the coordinate representation, is
suggested in Ref. \cite{MR96}, and its detailed formalism and
numerical solution for a particular case can be found in
Ref. \cite{ME98} and the
references therein. The basic ansatz of the RMF theory is a
Lagrangian density whereby nucleons are described as Dirac
particles which interact via the exchange of various mesons and
the photon. The mesons are the scalar sigma ($\sigma$),
vector omega ($\bf \omega$) and iso-vector vector rho ($\bf \vec
\rho$).  The rho ($\bf \vec \rho$) meson provides the necessary
isospin asymmetry. The scalar sigma meson moves in a
self-interacting field having cubic and quadratic terms with
strengths $g_2$ and $g_3$, respectively. The Lagrangian then
consists of the free baryon and meson parts and the interaction
part with minimal coupling, together with the nucleon mass $M$,
and $m_\sigma$, $g_\sigma$, $m_\omega$, $g_\omega$, $m_\rho$,
$g_\rho$  the masses and coupling constants of the respective
mesons and:
\begin{eqnarray}
\begin{array}{cc}
{\cal L} &= \bar \psi (i\rlap{/}\partial -M) \psi +
        \,{1\over2}\partial_\mu\sigma\partial^\mu\sigma-U(\sigma)
        -{1\over4}\Omega_{\mu\nu}\Omega^{\mu\nu} \\
        \ &+ {1\over2}m_\omega^2\omega_\mu\omega^\mu
        -{1\over4}{\vec R}_{\mu\nu}{\vec R}^{\mu\nu} +
        {1\over2}m_{\rho}^{2} \vec\rho_\mu\vec\rho^\mu
        -{1\over4}F_{\mu\nu}F^{\mu\nu} \\
        & -  g_{\sigma}\bar\psi \sigma \psi~
        -~g_{\omega}\bar\psi \rlap{/}\omega \psi~
        -~g_{\rho}  \bar\psi
        \rlap{/}\vec\rho
        \vec\tau \psi
        -~e \bar\psi \rlap{/}A \psi.
\label{Lagrangian}
\end{array}
\end{eqnarray}
The field tensors for the vector mesons are given as:
\begin{eqnarray}
\left\{
\begin{array}{lll}
   \Omega^{\mu\nu}   &=& \partial^\mu\omega^\nu-\partial^\nu\omega^\mu, \\
   {\vec R}^{\mu\nu} &=& \partial^\mu{\vec \rho}^\nu
                        -\partial^\nu{\vec \rho}^\mu
                        - g^{\rho} ( {\vec \rho}^\mu
                           \times {\vec \rho}^\nu ), \\
   F^{\mu\nu}        &=& \partial^\mu \vec A^\nu-\partial^\nu \vec A^\mu.
\end{array}   \right.
\label{tensors}
\end{eqnarray}
For a realistic description of nuclear properties, a nonlinear
self-coupling for the scalar mesons turns out crucial \cite{BB77}:
\begin{equation}
   U(\sigma)~=~\dfrac{1}{2} m^2_\sigma \sigma^2_{}
            ~+~\dfrac{g_2}{3}\sigma^3_{}~+~\dfrac{g_3}{4}\sigma^4_{}
\end{equation}
The classical variation principle gives the following equations of
motion :
\begin{equation}
   [ {\vec \alpha} \cdot {\vec p} +
     V_V ( {\vec r} ) + \beta ( M + V_S ( {\vec r} ) ) ]
     \psi_i ~=~ \epsilon_i\psi_i
\label{spinor1}
\end{equation}
for the nucleon spinors and
\begin{eqnarray}
\left\{
\begin{array}{lll}
  \left( -\Delta \sigma ~+~U'(\sigma) \right ) &=& g_\sigma\rho_s
\\
   \left( -\Delta~+~m_\omega^2\right )\omega^{\mu} &=&
                    g_\omega j^{\mu} ( {\vec r} )
\\
   \left( -\Delta~+~m_\rho^2\right) {\vec \rho}^{\mu}&=&
                    g_\rho \vec j^{\mu}( {\vec r} )
\\
          -\Delta~ A_0^{\mu} ( {\vec r} ) ~ &=&
                           e j_{\rho}^{\mu}( {\vec r} )
\end{array}  \right.
\label{mesonmotion}
\end{eqnarray}
for the mesons, where
\begin{eqnarray}
\left\{
\begin{array}{lll}
   V_V( {\vec r} ) &=&
      g_\omega\rlap{/}\omega + g_\rho\rlap{/}\vec\rho\vec\tau
         + \dfrac{1}{2}e(1-\tau_3)\rlap{\,/}\vec A , \\
   V_S( {\vec r} ) &=&
      g_\sigma \sigma( {\vec r} ) \\
\end{array}
\right. \label{vaspot}
\end{eqnarray}
are the vector and scalar potentials respectively and the source
terms for the mesons are
\begin{eqnarray}
\left\{
\begin{array}{lll}
   \rho_s &=& \sum_{i=1}^A \bar\psi_i \psi_i
\\
   j^{\mu} ( {\vec r} ) &=&
               \sum_{i=1}^A \bar \psi_i \gamma^{\mu} \psi_i
\\
   \vec j^{\mu}( {\vec r} ) &=&
          \sum_{i=1}^A \bar \psi_i \gamma^{\mu} \vec \tau  \psi_i
\\
   j^{\mu}_p ( {\vec r} ) &=&
      \sum_{i=1}^A \bar \psi_i \gamma^{\mu} \dfrac {1 - \tau_3} 2  \psi_i,
\end{array}  \right.
\label{mesonsource}
\end{eqnarray}
where the summations are over the valence nucleons only. It should
be noted that as usual, the present approach neglects the
contribution of negative energy states, i.e.,  no-sea
approximation, which means that the vacuum is not polarized. The
coupled equations Eq. (\ref{spinor1}) and Eq. (\ref{mesonmotion})
are nonlinear quantum field equations, and their exact solutions
are very complicated. Thus the mean field approximation is
generally used: i.e., the meson field operators in Eq.
(\ref{spinor1}) are replaced by their expectation values. Then the
nucleons move independently in the classical meson fields. The
coupled equations are self-consistently solved by iteration.
\par

For spherical nuclei, i.e., the systems with rotational symmetry,
the potential of the nucleon and the sources of meson fields
depend only on the radial coordinate $r$. The spinor is
characterized by the angular momentum quantum numbers $l$, $j$,
$m$, the isospin $t = \pm \dfrac 1 2$ for neutron and proton
respectively, and the other quantum number $i$. The Dirac spinor
has the form:
\begin{equation}
   \psi ( \vec r ) =
      \left( { {\mbox{i}  \dfrac {G_i^{lj}(r)} r {Y^l _{jm} (\theta,\phi)} }
      \atop
       { \dfrac {F_i^{lj}(r)} r (\vec\sigma \cdot \hat {\vec r} )
       {Y^l _{jm} (\theta,\phi)} } }
      \right) \chi _{t}(t),
\label{reppsi}
\end{equation}
where $Y^l _{jm} (\theta,\phi)$ are the spinor spherical harmonics
and $G_i^{lj}(r)$ and $F_i^{lj}(r)$ are the remaining radial wave
function for upper and lower components. They are normalized
according to
\begin{equation}
\int_0^{\infty}dr ( | G_i^{lj}(r) |^2 + | F_i^{lj}(r) |^2 ) = 1.
\end{equation}
The radial equation of spinor Eq. (\ref{spinor1}) can be reduced
as :
\begin{eqnarray}
\left\{
\begin{array}{lll}
   \epsilon_i G_i^{lj}(r) &=& ( - \dfrac {\partial} {\partial r}
      + \dfrac {\kappa_i} r )  F_i^{lj}(r) + ( M + V_S(r) + V_V(r) ) G_i^{lj}(r)
\\
   \epsilon_i F_i^{lj}(r) &=& ( + \dfrac {\partial} {\partial r}
      + \dfrac {\kappa_i} r )  G_i^{lj}(r)
      - ( M + V_S(r) - V_V(r) ) F_i^{lj}(r) .
\end{array}  \right.
\label{spinorradical}
\end{eqnarray}
where
\begin{displaymath}
   \kappa =
      \left\{
         \begin{array}{ll}
            -(j+1/2)  & for ~ j=l+1/2 \\
            +(j+1/2)  & for ~ j=l-1/2. \\
         \end{array}
      \right.
\end{displaymath}
The meson field equations become simply radical Laplace equations
of the form:
\begin{equation}
\left( \frac {\partial^2} {\partial r^2}  - \frac 2 r \frac
{\partial}   {\partial r} + m_{\phi}^2 \right)\phi = s_{\phi} (r),
\label{Ramesonmotion}
\end{equation}
$m_{\phi}$ are the meson masses for $\phi = \sigma, \omega,\rho$
and for photon ( $m_{\phi} = 0$ ). The source terms are:
\begin{eqnarray}
s_{\phi} (r) = \left\{
\begin{array}{ll}
-g_\sigma\rho_s - g_2 \sigma^2(r)  - g_3 \sigma^3(r)
& { \rm for ~ the ~  \sigma~  field } \\
g_\omega \rho_v    & {\rm for ~ the ~ \omega ~ field} \\
g_{\rho}  \rho_3(r)       & {\rm for~ the~ \rho~ field} \\
e \rho_c(r)  & {\rm for~ the~ Coulomb~ field}, \\
\end{array}
\right.
\end{eqnarray}

\begin{eqnarray}
\left\{
\begin{array}{lll}
   4\pi r^2 \rho_s (r) &=& \sum_{i=1}^A ( |G_i(r)|^2 - |F_i(r)|^2 ) \\
   4\pi r^2 \rho_v (r) &=& \sum_{i=1}^A ( |G_i(r)|^2 + |F_i(r)|^2 ) \\
   4\pi r^2 \rho_3 (r) &=& \sum_{p=1}^Z ( |G_p(r)|^2 + |F_p(r)|^2 )
               -  \sum_{n=1}^N ( |G_n(r)|^2 + |F_n(r)|^2 ) \\
   4\pi r^2 \rho_c (r) &=& \sum_{p=1}^Z ( |G_p(r)|^2 + |F_p(r)|^2 ) . \\
\end{array}
\right. \label{mesonsourceS}
\end{eqnarray}
The Laplace equation can in principle be solved by the Green
function:
\begin{equation}
\phi (r) = \int_0^{\infty} r'^2 dr' G_{\phi} (r,r') s_{\phi}(r'),
\end{equation}
where for massive fields
\begin{equation}
G_{\phi} (r,r') = \frac 1 {2m_{\phi}} \frac 1 {rr'} ( e^{-m_{\phi}
| r-r'|} -  e^{-m_{\phi} | r+r'|} )
\end{equation}
and for Coulomb field
\begin{equation}
G_{\phi} (r,r') = \left\{
\begin{array}{ll}
1/r  & {\rm for ~ r~ >~ r' } \\
1/r' & {\rm for ~r~ <~ r'} .\\
\end{array}
\right.
\end{equation}

Eqs. (\ref{spinorradical}) and (\ref{Ramesonmotion}) could be
solved self-consistently in the usual RMF approximation. For RMF,
however, as the classical meson fields are used, the equations of
motion for nucleons derived from Eq. (\ref{Lagrangian}) do not
contain pairing interaction. In order to have pairing interaction,
one has to quantize the meson fields which leads to a Hamiltonian
with two-body interaction. Following the standard procedure of
Bogoliubov transformation, a Dirac Hartree-Bogoliubov equation
could be derived and then a unified description of the mean field
and pairing correlation in nuclei could be achieved. For the
detail, see Ref. \cite{ME98} and the references therein. The RHB
equations are as following:
\begin{equation}
   \int d^3r'
   \left( \begin{array}{cc}
          h-\lambda &   \Delta \\
          \Delta    &  - h+\lambda
          \end{array} \right)
   \left( { \psi_U \atop\psi_V } \right)  ~
   = ~ E ~ \left( { \psi_U \atop \psi_V } \right),
\label{ghfb}
\end{equation}
where
\begin{equation}
   h(\vec r,\vec r') =  \left[ {\vec \alpha} \cdot {\vec p} +
      V_V ( {\vec r} ) + \beta ( M + V_S ( {\vec r} ) ) \right]
      \delta(\vec r,\vec r')
\label{NHamiltonian}
\end{equation}
is the Dirac Hamiltonian and the Fock term has been neglected as
is usually done in RMF. The pairing potential is :
\begin{eqnarray}
    \Delta_{kk'}(\vec r, \vec r')
    &=& \int d^3r_1 \int d^3r_1' \sum_{\tilde k \tilde k'}
    V_{kk',\tilde k \tilde k'} ( \vec r \vec r'; \vec r_1 \vec r_1' )
    \kappa_{\tilde k \tilde k'}  (\vec r_1, \vec r_1' ).
\label{gap}
\end{eqnarray}
It is obtained from one-meson exchange interaction $V_{kk',\tilde
k \tilde k'} (\vec r \vec r'; \vec r_1 \vec r_1' )$ in the
$pp$-channel and the pairing tensor $\kappa=V^*U^T$
\begin{eqnarray}
   \kappa_{k k'}( \vec r, \vec r') = < | a_{k} a _{k'} | >
   = \psi_V^{k}(\vec r) ^* \psi_U^{k'}(\vec r)^T
\end{eqnarray}
The nuclear density is as following:
\begin{eqnarray}
   \rho(\vec r ,\vec r' )
   = \sum_i^{lj}  \psi_V^{ilj} (\vec r) ^* \psi_V ^{ilj} (\vec r ' ) .
\end{eqnarray}
As in Ref. \cite{ME98}, $V_{kk',\tilde k \tilde k'}$ used for the
pairing potential in Eq. (\ref{gap}) is either the
density-dependent two-body force of zero range with the
interaction strength $V_0$ and the nuclear matter density
$\rho_0$:
\begin{equation}
   V(\mbox{\boldmath $r$}_1,\mbox{\boldmath $r$}_2) = V_0
     \delta(\mbox{\boldmath $r$}_1-\mbox{\boldmath $r$}_2)
     \frac{1}{4}\left[1-
     \mbox{\boldmath $\sigma$}_1\mbox{\boldmath $\sigma$}_2\right]
     \left(1 - \frac{\rho(r)}{\rho_0}\right)
\label{vpp}
\end{equation}
or Gogny-type finite range force with the parameter $\mu_i$,
$W_i$, $B_i$, $H_i$ and $M_i$ ($i=1,2$) as the finite range part
of the Gogny force \cite{BGG84}:
\begin{equation}
   V(\mbox{\boldmath $r$}_1,\mbox{\boldmath $r$}_2)
      ~=~\sum_{i=1,2}
       e^{((\mbox{\boldmath $r$}_1-\mbox{\boldmath$r$}_2) / \mu_i)^2}
       (W_i + B_i P^{\sigma} - H_i P^{\tau} - M_i P^{\sigma} P^{\tau})
\label{vpp2}
\end{equation}
A Lagrange multiplier $\lambda$ is introduced to fix the particle
number for the neutron and proton as $N = \mbox {Tr} \rho _n$ and
$Z = \mbox {Tr} \rho _p$.

In order to describe both the continuum and the bound states
self-consistently, the RHB theory must be solved in coordinate
representation, i.e., the. Relativistic Continuum
Hartree-Bogoliubov ( RCHB ) theory \cite{ME98}. It is then
applicable to both exotic nuclei and normal nuclei. In Eq.
(\ref{ghfb}), the spectrum of the system is unbound from above and
from below the Fermi surface, and the eigenstates occur in pairs
of opposite energies. When spherical symmetry is imposed on the
solution of the RCHB equations, the wave function can be
conveniently written as
\begin{equation}
   \psi^i_U =
      \left( {\displaystyle {\mbox{i} \frac {G_U^{ilj}(r)} r }  \atop
     {\displaystyle \frac {F_U^{ilj}(r)} r
        (\vec\sigma \cdot \hat {\vec r} )  } }
              \right) {Y^l _{jm} (\theta,\phi)}  \chi_{t}(t) ,
   \psi^i_V =
       \left( {\displaystyle {\mbox{i} \frac {G_V^{ilj}(r)} r }  \atop
          {\displaystyle \frac {F_V^{ilj}(r)} r
             (\vec\sigma \cdot \hat {\vec r} )
       } } \right)  {Y^l _{jm} (\theta,\phi)}  \chi_{t}(t).
\end{equation}

The above equation Eq. (\ref{ghfb})  depends only on the radial
coordinates and can be expressed as the following
integro-differential equation:
\begin{eqnarray}
\left\{
   \begin{array}{lll}
      \displaystyle
      \frac {d G_U(r)} {dr} + \frac {\kappa} r G_U(r) -
       ( E + \lambda-V_V(r) + V_S(r) ) F_U(r) +
         r \int r'dr' \Delta(r,r')  F_V(r') &=& 0  \\
      \displaystyle
      \frac {d F_U(r)} {dr} - \frac {\kappa} r F_U(r) +
       ( E + \lambda-V_V(r)-V_S(r) ) G_U(r) +
         r \int r'dr' \Delta(r,r') G_V(r') &=& 0 \\
      \displaystyle
      \frac {d G_V(r)} {dr} + \frac {\kappa} r G_V(r) +
       ( E - \lambda+V_V(r)-V_S(r) ) F_V(r) +
         r \int r'dr' \Delta(r,r') F_U(r') &=& 0 \\
      \displaystyle
      \frac {d F_V(r)} {dr} - \frac {\kappa} r F_V(r) -
       ( E - \lambda+V_V(r)+V_S(r) ) G_V(r) +
         r \int r'dr' \Delta(r,r')  G_U(r') &=& 0, \\
\end{array}
\right. \label{CoupEq}
\end{eqnarray}
where the nucleon mass is included in the scalar potential
$V_S(r)$. Eq. (\ref{CoupEq}), in the case of $\delta$-force given
in Eq. (\ref{vpp}), is reduced to normal coupled differential
equations and can be solved with shooting method by Runge-Kutta
algorithms. For the case of Gogny force, the coupled
integro-differential equations are discretized in the space and
solved by the finite element methods, see Ref. \cite{ME98}.
Instead of solving Eqs. (\ref{spinorradical}) and
(\ref{Ramesonmotion}) self-consistently for the RMF case, now we
have to solve Eqs. ({\ref{CoupEq}) and (\ref{Ramesonmotion})
self-consistently for the RCHB case. As the calculation for Gogny
force is very time-consuming, we use them for one nucleus and fix
the interaction strength of the $\delta$-force given in Eq.
(\ref{vpp}).

\section{Results and discussions}

\subsection{Binding energies and two neutron separation energies}

The numerical techniques of the RCHB theory can be found in
Ref.\cite{ME98} and the references therein. In the present
calculations, we follow the procedures in Ref.\cite{ME98} and
solve the RCHB equations in a box with the size $R=20$ fm and a
step size of 0.1 fm. The parameter set NL-SH \cite{SNR93} is used,
which aims at describing both the stable and exotic nuclei. The
use of the TM1 parameter set provides similar results and we show
here only those of the NL-SH parameter \cite{ST94}. The density
dependent $\delta$-force in the pairing channel with
$\rho_0=0.152$ fm$^{-3}$ is used and its strength $V_0=650$ MeV
$\cdot$ fm$^{-3}$ is fixed by the Gogny force as in
Ref.\cite{ME98}. The contribution from continua is restricted
within a cut-off energy $E_{cut}\sim 120$MeV.

In this work the RCHB calculation is restricted to the spherical
shape, which is a good approximation for most proton magic nuclei
with $Z=$ 8, 20, 28, 50, 82. The RCHB code was also carried out
for the Zr isotopes, with the considerations of the sub-magic
number $Z=40$ and the fact that most of the investigations with
non-relativistic codes and with relativistic RMF+BCS codes show
that the nuclei above $^{122}$Zr are spherical \cite{MR98}.

We calculated ground state properties of all the even-even O, Ca,
Ni, Zr, Sn, Pb isotopes ranging from the proton drip-line to the
neutron drip-line with the RCHB code. We here present the binding
energy $E_b$ calculated from the RCHB method for these isotope
chains and the corresponding data available \cite{AW95} in the
last two columns in table I-IV. We do not include the tables for
the Sn and Pb isotopes here to save the space.  The difference
($\Delta E_b$) between the experimental and calculated binding
energies is less than 3.5 MeV for most nuclei, which is less than
$1\%$ of the experimental values. The large difference ($\Delta
E_b\sim 10$ MeV) for some Zr isotopes are found, which are mainly
due to deformation. The two neutron separation energy $S_{2n}$
defined as
\begin{equation}
 S_{2n}=E_b(Z,A+2)-E_b(Z,A)
\end{equation}
is quite a sensitive quantity to test a microscopic theory. The
two neutron separation energy becomes negative when the nucleus
becomes unstable against the two neutron emission.  Hence, the
drip line nucleus for a corresponding isotope chain is the one
before the nucleus with the negative $S_{2n}$.  In Fig. \ref{fig.
S2n} both the theoretical and the available experimental $S_{2n}$
are presented as a function of neutron number $N$ for the O, Ca,
Ni, Zr, Sn, Pb isotope chains, respectively. The good coincidence
between experiment and calculation is clearly seen.

\begin{figure}
\centerline{\psfig{figure=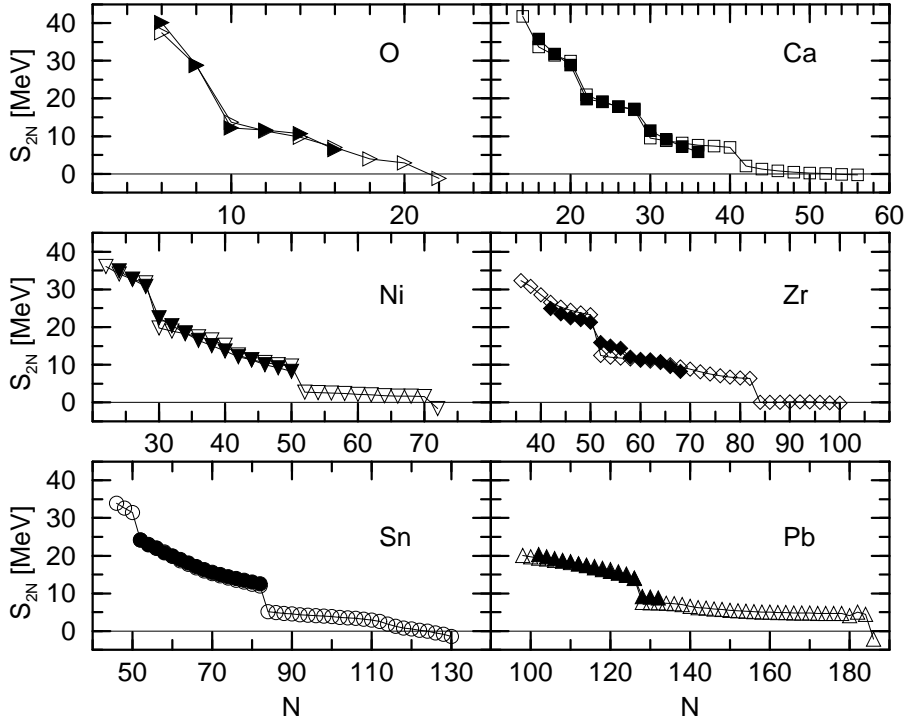,width=12.cm}}
 \caption{The two neutron separation energies $S_{2n}$ for the proton
 magic isotopes are plotted as a function of the neutron number.
 The figures from left to right and from up to down are those for O, Ca,
 Ni, Zr, Sn and Pb, respectively.  The experimental values are
 denoted by the solid symbols and the calculated ones by the open symbols.}
 \label{fig. S2n}
\end{figure}

\begin{table}[htbp]
\caption{The ground state properties of even O isotopes calculated
with the RCHB theory with the NL-SH parameter. Listed are
neutron($r_n$), proton($r_p$), matter($r_m$), and charge($r_c$)
root mean square radii as well as binding energies $E_b$. The
corresponding experimental values $r_c$ and $E_b$ are also
included when available.}
\begin{center}
\begin{tabular}{c@{\hspace{1ex}}|@{\hspace{1ex}}ccccc@{\hspace{2ex}}|@{\hspace{2ex}}cc}
\hline\hline
 Nucleus & \multicolumn{4}{c}{\hspace{8ex} Radius (fm)} &
        & \multicolumn{2}{c}{$E_b$ (MeV)}   \\
 $^A$O & $r_c^{\rm exp.}$ & $r_n$ & $r_p$ &
$r_m$ & $r_c$ & Exp.  &  RCHB \\ \hline
$^{12}$O & & 2.238  &   2.824  &   2.643 & 2.935 & -58.549 & -62.169 \\
$^{14}$O & & 2.361  &   2.602  &   2.501 & 2.722 & -98.733 &  -99.641 \\
$^{16}$O & 2.693 & 2.551  &   2.578  &   2.565 & 2.699 & -127.619 & -128.347 \\
$^{18}$O & 2.727 & 2.738  &   2.572  &   2.665 & 2.694 & -139.807 & -142.081 \\
$^{20}$O & & 2.885  &   2.575  &   2.765 & 2.696 & -151.370 & -153.397 \\
$^{22}$O & & 3.012  &   2.580  &   2.862 & 2.701 & -162.030 & -163.124 \\
$^{24}$O & & 3.232  &   2.607  &   3.038&  2.727 & -168.500 & -170.169\\
$^{26}$O & & 3.375  &   2.656  &   3.171 & 2.774 &  & -174.126 \\
$^{28}$O & & 3.503  &   2.705  &   3.295 & 2.821 &  & -177.040\\
$^{30}$O & &  &  &    &  &  & -174.453 \\
\hline\hline
\end{tabular}
\end{center}
\end{table}

\begin{table}[htbp]
\caption{The same as Tab. I, but for even Ca isotopes.}
\vspace{1ex}
\begin{center}
\begin{tabular}{c@{\hspace{1ex}}|@{\hspace{1ex}}ccccc@{\hspace{2ex}}|@{\hspace{2ex}}cc}
\hline\hline
 Nucleus & \multicolumn{4}{c}{\hspace{8ex} Radius (fm)} &
        & \multicolumn{2}{c}{$E_b$ (MeV)}   \\
$^A$Ca & $r_c^{\rm exp.}$ & $r_n$ & $r_p$ & $r_m$ & $r_c$ & Exp. & RCHB \\
\hline
$^{34}$Ca & & 3.015 & 3.368 & 3.227 & 3.462 & -245.600 & -244.911 \\
$^{36}$Ca & & 3.133 & 3.355 & 3.258 & 3.449 & -281.360 & -278.557 \\
$^{38}$Ca & & 3.229 & 3.354 & 3.295 & 3.448 & -313.122 & -310.051 \\
$^{40}$Ca & 3.478 & 3.311 & 3.359 & 3.335 & 3.452 & -342.052 & -340.006 \\
$^{42}$Ca & 3.508 & 3.391 & 3.358 & 3.375 & 3.452 & -361.895 & -360.995 \\
$^{44}$Ca & 3.518 & 3.463 & 3.360 & 3.416 & 3.454 & -380.960 & -380.086 \\
$^{46}$Ca & 3.498 & 3.527 & 3.364 & 3.457 & 3.458 & -398.769 & -397.937 \\
$^{48}$Ca & 3.479 & 3.584 & 3.369 & 3.496 & 3.463 & -415.991 & -414.910 \\
$^{50}$Ca & & 3.703 & 3.394 & 3.583 & 3.487 & -427.491 & -424.443 \\
$^{52}$Ca & & 3.808 & 3.419 & 3.664 & 3.512 & -436.600 & -433.223 \\
$^{54}$Ca & & 3.899 & 3.447 & 3.738 & 3.538 & -443.800 & -441.354 \\
$^{56}$Ca & & 3.980 & 3.475 & 3.807 & 3.566 & -449.600 & -448.998 \\
$^{58}$Ca & & 4.054 & 3.503 & 3.873 & 3.593 & & -456.304 \\
$^{60}$Ca & & 4.125 & 3.532 & 3.937 & 3.621 & & -463.260 \\
$^{62}$Ca & & 4.182 & 3.552 & 3.989 & 3.641 & & -465.323 \\
$^{64}$Ca & & 4.244 & 3.572 & 4.046 & 3.660 & & -466.562 \\
$^{66}$Ca & & 4.314 & 3.591 & 4.108 & 3.679 & & -467.320 \\
$^{68}$Ca & & 4.400 & 3.608 & 4.183 & 3.696 & & -467.752 \\
$^{70}$Ca & & 4.513 & 3.622 & 4.277 & 3.710 & & -467.957 \\
$^{72}$Ca & & 4.636 & 3.634 & 4.381 & 3.721 & & -467.993 \\
$^{74}$Ca & &  &  &  &  & & -467.645 \\
\hline\hline
\end{tabular}
\end{center}
\end{table}

\begin{table}[htbp]
\caption{The same as Tab. I, but for even Ni isotopes.}
\vspace{1ex}
\begin{center}
\begin{tabular}{c@{\hspace{1ex}}|@{\hspace{1ex}}ccccc@{\hspace{2ex}}|@{\hspace{2ex}}cc}
\hline\hline
 Nucleus & \multicolumn{4}{c}{\hspace{8ex} Radius (fm)} &
        & \multicolumn{2}{c}{$E_b$ (MeV)}   \\
$^A$Ni & $r_c^{\rm exp.}$ & $r_n$ & $r_p$ & $r_m$ & $r_c$ & Exp.
&  RCHB \\ \hline
$^{48}$Ni & & 3.344 & 3.669 & 3.537 & 3.755 & & -348.814 \\
$^{50}$Ni & & 3.414 & 3.651 & 3.549 & 3.738 & -385.500 & -384.925 \\
$^{52}$Ni & & 3.476 & 3.641 & 3.566 & 3.727 & -420.460 & -419.044 \\
$^{54}$Ni & & 3.532 & 3.634 & 3.585 & 3.721 & -453.150 & -451.786 \\
$^{56}$Ni & & 3.582 & 3.630 & 3.606 & 3.717 & -483.988 & -483.498 \\
$^{58}$Ni & 3.776 & 3.668 & 3.658 & 3.663 & 3.745 & -506.454 & -503.244 \\
$^{60}$Ni & 3.813 & 3.745 & 3.686 & 3.718 & 3.772 & -526.842 & -522.146 \\
$^{62}$Ni & 3.842 & 3.817 & 3.712 & 3.770 & 3.797 & -545.259 & -540.336 \\
$^{64}$Ni & 3.860 & 3.886 & 3.736 & 3.821 & 3.821 & -561.755 & -557.802 \\
$^{66}$Ni & & 3.954 & 3.758 & 3.872 & 3.842 & -576.830 & -574.370 \\
$^{68}$Ni & & 4.025 & 3.778 & 3.925 & 3.862 & -590.430 & -589.508 \\
$^{70}$Ni & & 4.075 & 3.794 & 3.965 & 3.877 & -602.600 & -602.094 \\
$^{72}$Ni & & 4.122 & 3.809 & 4.003 & 3.893 & -613.900 & -613.412 \\
$^{74}$Ni & & 4.168 & 3.826 & 4.042 & 3.908 & -623.900 & -623.982 \\
$^{76}$Ni & & 4.211 & 3.842 & 4.079 & 3.924 & -633.100 & -634.009 \\
$^{78}$Ni & & 4.252 & 3.858 & 4.115 & 3.940 & -641.400 & -643.615 \\
$^{80}$Ni & & 4.357 & 3.873 & 4.194 & 3.955 & & -646.356 \\
$^{82}$Ni & & 4.457 & 3.888 & 4.271 & 3.969 & & -648.979 \\
$^{84}$Ni & & 4.552 & 3.902 & 4.346 & 3.983 & & -651.488 \\
$^{86}$Ni & & 4.639 & 3.917 & 4.417 & 3.998 & & -653.855 \\
$^{88}$Ni & & 4.717 & 3.934 & 4.482 & 4.014 & & -656.036 \\
$^{90}$Ni & & 4.782 & 3.954 & 4.540 & 4.035 & & -658.026 \\
$^{92}$Ni & & 4.835 & 3.979 & 4.592 & 4.058 & & -659.852 \\
$^{94}$Ni & & 4.882 & 4.006 & 4.638 & 4.085 & & -661.570 \\
$^{96}$Ni & & 4.925 & 4.034 & 4.683 & 4.113 & & -663.220 \\
$^{98}$Ni & & 4.967 & 4.064 & 4.727 & 4.142 & & -664.767 \\
$^{100}$Ni & &  &  &  &  & & -663.123 \\
\hline\hline
\end{tabular}
\end{center}
\end{table}

\begin{table}[htbp]
\vspace{0.2cm} \caption{The same as Tab. I, but for even Zr
isotopes.}
\begin{center}
\begin{tabular}{c@{\hspace{1ex}}|@{\hspace{1ex}}ccccc@{\hspace{2ex}}|@{\hspace{2ex}}cc}
\hline\hline
 Nucleus & \multicolumn{4}{c}{\hspace{8ex} Radius (fm)} &
        & \multicolumn{2}{c}{$E_b$ (MeV)}   \\
$^A$Zr & $r_c^{\rm exp.}$ & $r_n$ & $r_p$ & $r_m$ & $r_c$ & Exp.
&  RCHB \\ \hline
$^{74}$Zr & & 3.932 & 4.149 & 4.051 & 4.225 & & -568.120 \\
$^{76}$Zr & & 3.984 & 4.150 & 4.072 & 4.226 & & -600.450 \\
$^{78}$Zr & & 4.035 & 4.155 & 4.097 & 4.231 & & -631.267 \\
$^{80}$Zr & & 4.088 & 4.163 & 4.126 & 4.239 & -669.800 & -659.843 \\
$^{82}$Zr & & 4.133 & 4.166 & 4.149 & 4.242 & -694.700 & -686.399 \\
$^{84}$Zr & & 4.174 & 4.168 & 4.171 & 4.244 & -718.190 & -711.679 \\
$^{86}$Zr & & 4.214 & 4.171 & 4.194 & 4.247 & -740.640 & -736.106 \\
$^{88}$Zr & & 4.251 & 4.176 & 4.217 & 4.252 & -762.606 & -759.890 \\
$^{90}$Zr & 4.270 & 4.286 & 4.181 & 4.240 & 4.257 & -783.893 & -783.173 \\
$^{92}$Zr & 4.305 & 4.346 & 4.204 & 4.285 & 4.280 & -799.722 & -795.603 \\
$^{94}$Zr & 4.330 & 4.403 & 4.226 & 4.329 & 4.302 & -814.677 & -807.665 \\
$^{96}$Zr & 4.349 & 4.458 & 4.248 & 4.372 & 4.323 & -828.994 & -819.421 \\
$^{98}$Zr & &  4.511 &  4.269 &  4.414 &  4.343  & -840.972 & -830.903 \\
$^{100}$Zr & & 4.563 &  4.288 &  4.455 &  4.362  & -852.440 & -842.122 \\
$^{102}$Zr & & 4.613 &  4.306 &  4.496 &  4.380  & -863.720 & -853.053 \\
$^{104}$Zr & & 4.664 &  4.323 &  4.536 &  4.397  & -874.500 & -863.613 \\
$^{106}$Zr & & 4.716 &  4.339 &  4.578 &  4.413  & -884.000 & -873.647 \\
$^{108}$Zr & & 4.769 &  4.355 &  4.620 &  4.428  & -892.300 & -883.105 \\
$^{110}$Zr & & 4.817 &  4.370 &  4.660 &  4.443  & & -891.983 \\
$^{112}$Zr & & 4.856 &  4.385 &  4.693 &  4.458  & & -900.136 \\
$^{114}$Zr & & 4.889 &  4.401 &  4.723 &  4.473  & & -907.646 \\
$^{116}$Zr & & 4.920 &  4.417 &  4.753 &  4.488  & & -914.714 \\
$^{118}$Zr & & 4.951 &  4.433 &  4.782 &  4.504  & & -921.459 \\
$^{120}$Zr & & 4.981 &  4.449 &  4.810 &  4.520  & & -927.963 \\
$^{122}$Zr & & 5.010 &  4.466 &  4.838 &  4.537  & & -934.285 \\
$^{124}$Zr & & 5.103 &  4.474 &  4.909 &  4.545  & & -934.256 \\
$^{126}$Zr & & 5.192 &  4.483 &  4.977 &  4.553  & & -934.271 \\
$^{128}$Zr & & 5.275 &  4.491 &  5.043 &  4.562  & & -934.329 \\
$^{130}$Zr & & 5.354 &  4.499 &  5.106 &  4.570  & & -934.419 \\
$^{132}$Zr & & 5.429 &  4.508 &  5.167 &  4.578  & & -934.526 \\
$^{134}$Zr & & 5.501 &  4.516 &  5.226 &  4.587  & & -934.618 \\
$^{136}$Zr & & 5.563 &  4.527 &  5.280 &  4.597  & & -934.640 \\
$^{138}$Zr & &  &  &  &  & & -934.540 \\
\hline\hline
\end{tabular}
\end{center}
\end{table}

We See from Fig. \ref{fig. S2n} and Tab. I-IV the proton magic
even mass nuclei at the neutron drip-line are predicted as
$^{28}$O, $^{72}$Ca, $^{98}$Ni, $^{136}$Zr, $^{174}$Sn, and
$^{266}$Pb, respectively. Out of these neutron drip line nuclei,
we have experimental information only for the O isotope.  The
experimental efforts were made to investigate $^{26}$O
\cite{FA96,GU90} and $^{28}$O \cite{TA97} whether they are bound.
These nuclei were found unstable and the heaviest O isotope was
concluded to be $^{24}$O. Hence, the present calculation is not
successful in reproducing the O neutron drip line.  Numerous
theoretical studies of binding energies of the neutron-rich O
nuclei have been made \cite{SLR99}, but these calculations failed
to reproduce that $^{24}$O is the drip line nucleus as the case of
RCHB with NL-SH presented here. It is known that in the
relativistic mean field theory, one parameter set is used to
describe all nuclei in the nuclear chart, which is quite a
challenge. In fact, it is well known that the mean field theory is
difficult for light nuclei. Therefore we have NL1, NL-SH, NL3, and
TM1 for the heavy system and NL2 and TM2 for the light system. The
heavy oxygen isotopes lie just at the border line between the
light and heavy groups. Therefore it is not surprising that the
NL-SH, NL1, NL2, NL3, TM1, TM2 and the non-relativistic
counterparts mostly do not predict correctly the neutron drip line
which is already known experimentally to be at $^{24}$O.

One may suspect that the drip-line nucleus for Ca is predicted to
be $^{70}$Ca instead of $^{72}$Ca. Most calculations predict the
Ca isotopes with $N>50$ would be unbound, as the more neutrons
should fill the continuum region above the $N=50$ shell. For
example, in Ref.\cite{FTZ00} and Ref. \cite{IM00}, the drip-line
nucleus is predicted as $^{70}$Ca with the HFB and Skyrme HF
method, respectively.

The reason for the unexpected bound nuclei $^{72}$Ca is caused by
the halo effect, which also results in the disappearance of the
normal $N=50$ magic number in the RCHB calculation. Seen along the
$S_{2n}$ to $N$ curve in Fig. \ref{fig. S2n}, strong kinks can be
clearly seen at $N=8$, $N=20, 28, 40$, $N=28,50$, $N=50,82$ and
$N=126$ for O, Ca, Ni, Zr, Sn, Pb isotope chains, respectively.
All these numbers correspond to the neutron magic numbers, which
come from the large gaps of single particle energy levels.
However, no kink appears at the $N=50$ magic number for Ca
isotopes in Fig. \ref{fig. S2n}. In stable nuclei, the $N=50$
magic number is given by the big energy gap between the $1g_{9/2}$
and $3s,~2d$ levels. However in the vicinity of the drip-line
region of Ca isotopes, the single particle energy of $3s_{1/2}$
would decrease with the halo effect and its zero centrifugal
potential barrier, the large gap between $1g_{9/2}$ and $3s_{1/2}$
disappears. Therefore, the nucleus $^{70}$Ca is no more a
double-magic nucleus and furthermore $^{72}$Ca is also bound. This
disappearance of the $N=50$ magic number at the neutron drip line
is due to the halo property of the neutron density with the
spherical shape being kept, different from the disappearance of
the $N=20$ magic number due to deformation (the unbound $^{26,
28}$O mentioned above). Most recently a new magic number $N=16$
has been discovered in neutron drip-line light-nuclei region which
is also considered as owing to the halo effect \cite{OZ00}. It can
be expected also that near drip line, more disappearance of the
traditional magic numbers and regeneration of new magic number
would be found with the same mechanism.

Another remarkable phenomenon in Fig. \ref{fig. S2n} is that the
$S_{2n}$ values for exotic Ca isotopes near neutron drip line are
very close to zero in the large mass region, i.e., $S_{2n}\approx
2.06,~1.24,~0.76,~0.43,~0.21,~0.04$ MeV for
$A=62,~64,~66,~68,~70,~72$ isotopes, respectively. If one regards
$^{60}$Ca as a core, then the several valence neutrons occupying
levels above the $N=40$ sub-shell for these $A>60$ exotic nuclei
are all weakly bound and can be scattered easily to the continuum
levels due to the pairing interaction, especially for
$^{66-72}$Ca. This case is very similar to $S_{2n}$ in Zr isotopes
with $N>82$ (see Fig. \ref{fig. S2n}) \cite{MR98} . Note that for
Sn isotopes in the vicinity of the drip line \cite{MT99}, however,
$S_{2n}$ decrease rather fast with the mass number and for Ni
isotopes \cite{ME98}, $S_{2n}$ are quite large ($\sim 2 $MeV) and
undergo a sudden drop to negative value at the drip-line nucleus.
Such behavior of $S_{2n}$ in Ca isotopes with $N>40$ gives us a
hint that there exist giant halos in these nuclei, just as what
happens in Zr isotopes with $N>82$ \cite{MR98}.

\subsection{Nuclear root mean square radii}

We discuss here the nuclear radii, which are the important basic
physical quantity to describe atomic nuclei as well as the nuclear
binding energies.  In the mean field theory, the root mean square
(rms) proton, neutron, mass radii ($r_p,~r_n,~r_m$) can be
directly deduced from their density distributions $\rho$:
\begin{equation}
r_{p(n)}=<r_{p(n)}^2>^{1/2}=(\dfrac{\int \rho_{p(n)} r^2
d\tau}{\int \rho_{p(n)} d\tau})^{1/2},
\end{equation}
and
\begin{equation}
r_{m}=<r_{m}^2>^{1/2}=(\dfrac{\int (\rho_{p}+\rho_{n}) r^2
d\tau}{\int (\rho_{p}+\rho_{n}) d\tau})^{1/2}.
\end{equation}
The theoretical rms charge radii $r_c$ are connected with the
proton radii $r_p$ as:
\begin{equation}
r_c^2=r_p^2+0.64~ {\rm fm}^2,
\end{equation}
when the small isospin dependence is ignored for exotic nuclei
\cite{DNW96}.

\begin{figure}
\centerline{\epsfig{figure=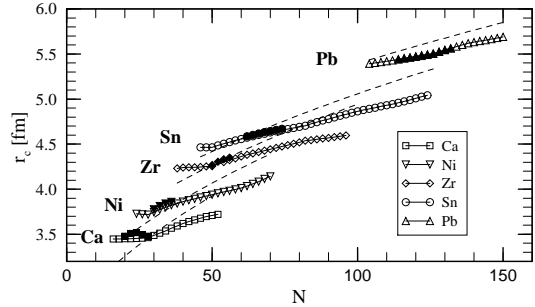,width=7.cm}} \caption{The root
mean square charge radii for the proton magic even mass nuclei are
plotted as a function of the neutron number.  The dotted curves
correspond to the radii calculated with the $A^{1/3}$ formula.}
\label{fig. r_ch}
\end{figure}

\begin{figure}
\centerline{\epsfig{figure=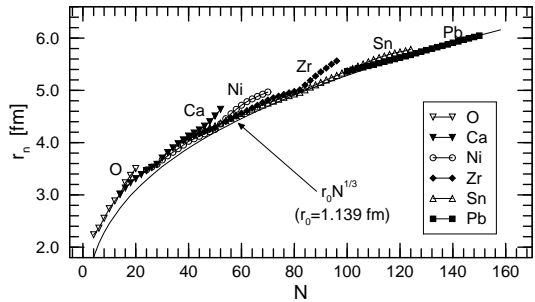,width=7.cm}} \caption{The root
mean square neutron radii for the proton magic even mass nuclei
are plotted as a function of the neutron number.  The solid curve
corresponds to the radii calculated with the $N^{1/3}$ formula.}
\label{fig. r_n}
\end{figure}

\begin{figure}
\centerline{\epsfig{figure=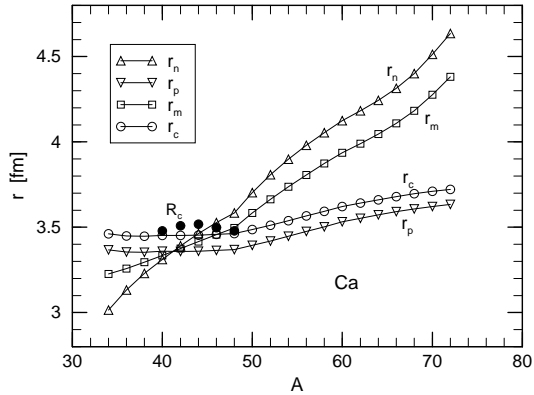,width=7.cm}} \caption{The
root mean square radii for the proton, neutron, charge and matter
distributions are plotted as a function of the neutron number for
the Ca isotope.  The available experimental values are denoted by
solid symbols.} \label{fig. radii}
\end{figure}

It is interesting to investigate the neutron number dependence of
the proton radii (or charge radii) and the neutron radii for these
proton-magic isotope chains. In Fig. \ref{fig. r_ch}, we show the
rms charge radii $r_c$ obtained from the RCHB theory (open
symbols) and the data available (solid symbols) for the even-even
Ca, Ni, Zr, Sn and Pb isotopes. The values for the O isotopes are
not plotted here considering the systematic deviation of the mean
field theory for light isotopes as discussed above. As it could be
seen, the RCHB calculations reproduce the data very well (within
1.5$\%$). For a given isotopic chain, an approximate linear $N$
dependence of the calculated rms charge radii $ r_c$ is clearly
seen. However, the variation of $r_c$ for a given isotopic chain
deviates from the general accepted simple $A^{1/3}$ law (denoted
by dashed lines in Fig. \ref{fig. r_ch}), which shows a strong
isospin dependence of nuclear charge radii is necessary for nuclei
with extreme $N/Z$ ratio. Recently, we have investigated these
charge radii as well as the large amount of experimental data and
extract a formula with the $Z^{1/3}$ dependence and isospin
dependence to better describe these charge radii all over the
nuclear isotope chart \cite{ZH02}.

We also plot in Fig. \ref{fig. r_n} the neutron radii $r_n$ from
the RCHB calculation for the even-even nuclei of the O, Ca, Ni,
Zr, Sn and Pb isotope chains. The prediction $r_n$ curve using the
simple experiential equation $r_n=r_0\cdot N^{1/3}$ with
$r_0=1.139$ normalizing to $^{208}$Pb is also represented in the
figure. Except for the O and Ca isotopes, it is very interesting
to see that the simple formula for $r_n$ agrees with the
calculated neutron radii except for two anomalies.  One appears in
the Ca chain above $N=40$, and the other in Zr chain above $N=82$.
The increase of exotic Ni and Sn nuclei is not as rapid as that in
Ca and Zr chains. The regions of the abnormal increases of the
neutron radii are just the same as those for $S_{2n}$.  Both
behaviors are connected with the formation of giant halo. As
Calcium is a light element, the giant halos in exotic Ca isotopes
would be much more easily accessed experimentally than in the Zr
chain.

In Tab. II we present the calculated neutron, proton, matter, and
charge radii for all even Ca isotopes, as well as several
available charge radii data. Those radii against mass number $A$
are plotted in Fig. \ref{fig. radii}. For nuclear charge radii
$r_c$, the well-known parabolic behavior along $^{40-48}$Ca is not
reproduced in our calculation. It is mainly due to the improper
spherical supposition to the real deformed $^{42, 44, 46}$Ca
nuclei. It can be clearly seen that the radii $r_p$, $r_n$, $r_m$
and $r_c$ all increase with the mass number $A$. The mass radii
$r_m$ as well as $r_n$ increase much faster than $r_p$ and $r_c$.
Besides the normal increase of $r_n$ and $r_m$ with $A$, a
slightly up-bend tendency occurs in the neutron drip line region
from $A= 62$ up to $A=72$.

\subsection{Density distribution and halos}

We shall concentrate from here the giant halo properties for the
Ca isotopes as mainly due to the easier access by the present
experimental techniques.  For exotic nuclei, physics connected
with the low density region in the tails of the neutron and proton
distributions have attracted a lot of attention in nuclear physics
as well as in other fields such as astrophysics. It is therefore
of great importance to look into the matter distribution and see
how the densities change with the proton to neutron ratio in these
nuclei. As the density here is obtained from a fully microscopic
and parameter free model which is well supported by the
experimental binding energies, we now proceed to examine the
density distributions of the whole chain of Calcium isotopes in
this section and study the relation between the development of
halo and shell effects within the model.

\begin{figure}[t]
\centerline{\epsfig{figure=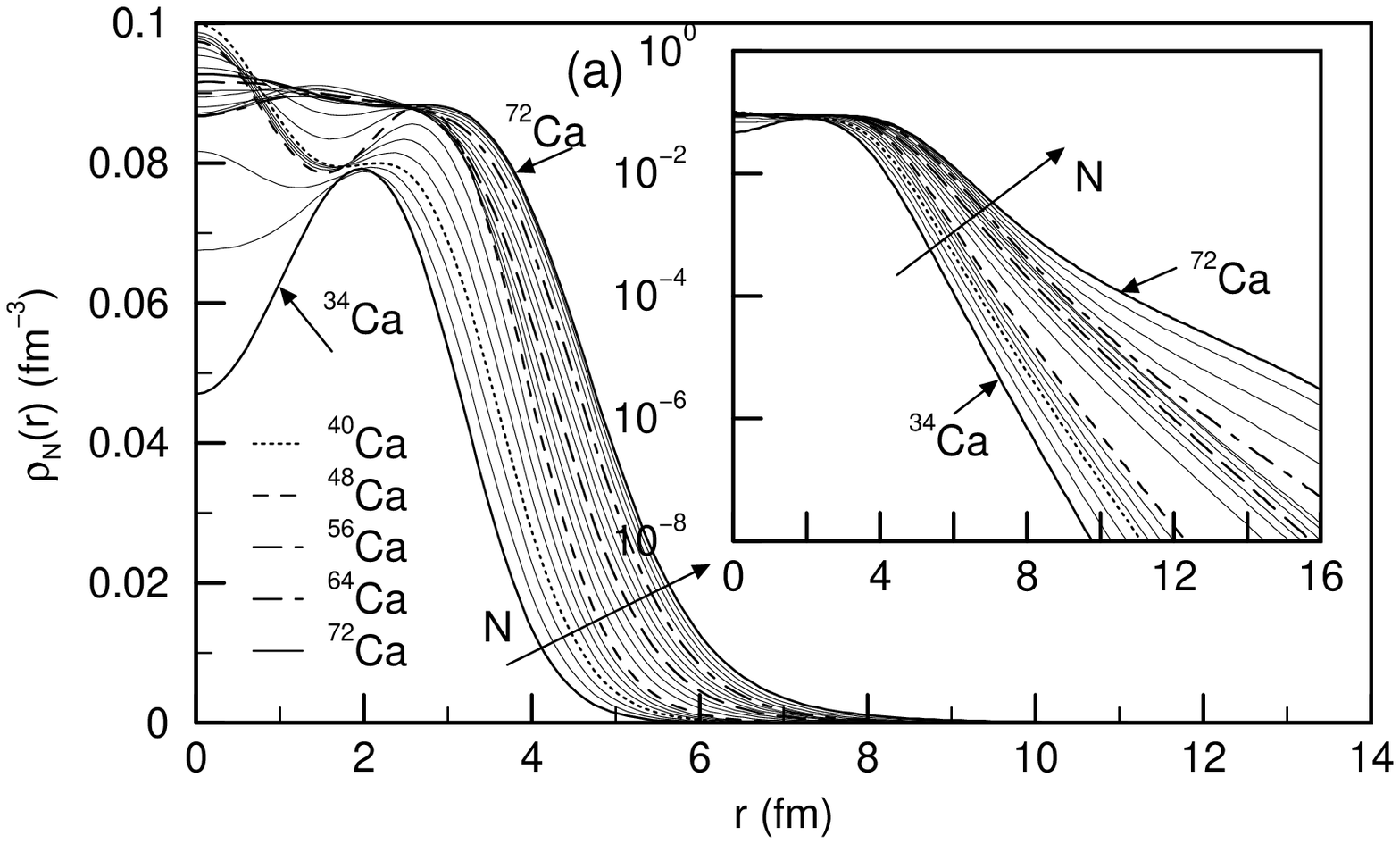,width=7.cm}}

\centerline{\epsfig{figure=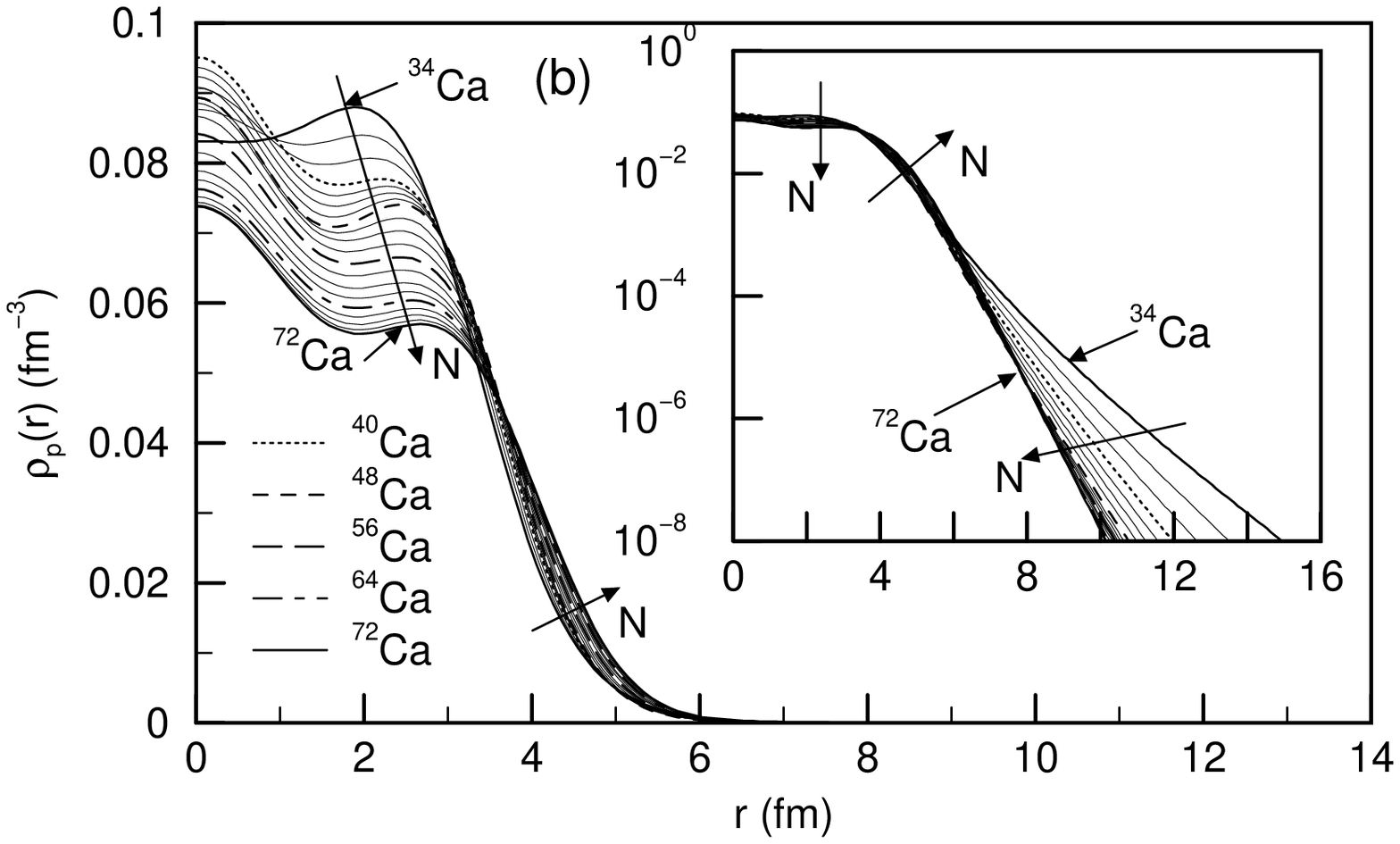,width=7.cm}}

\centerline{\epsfig{figure=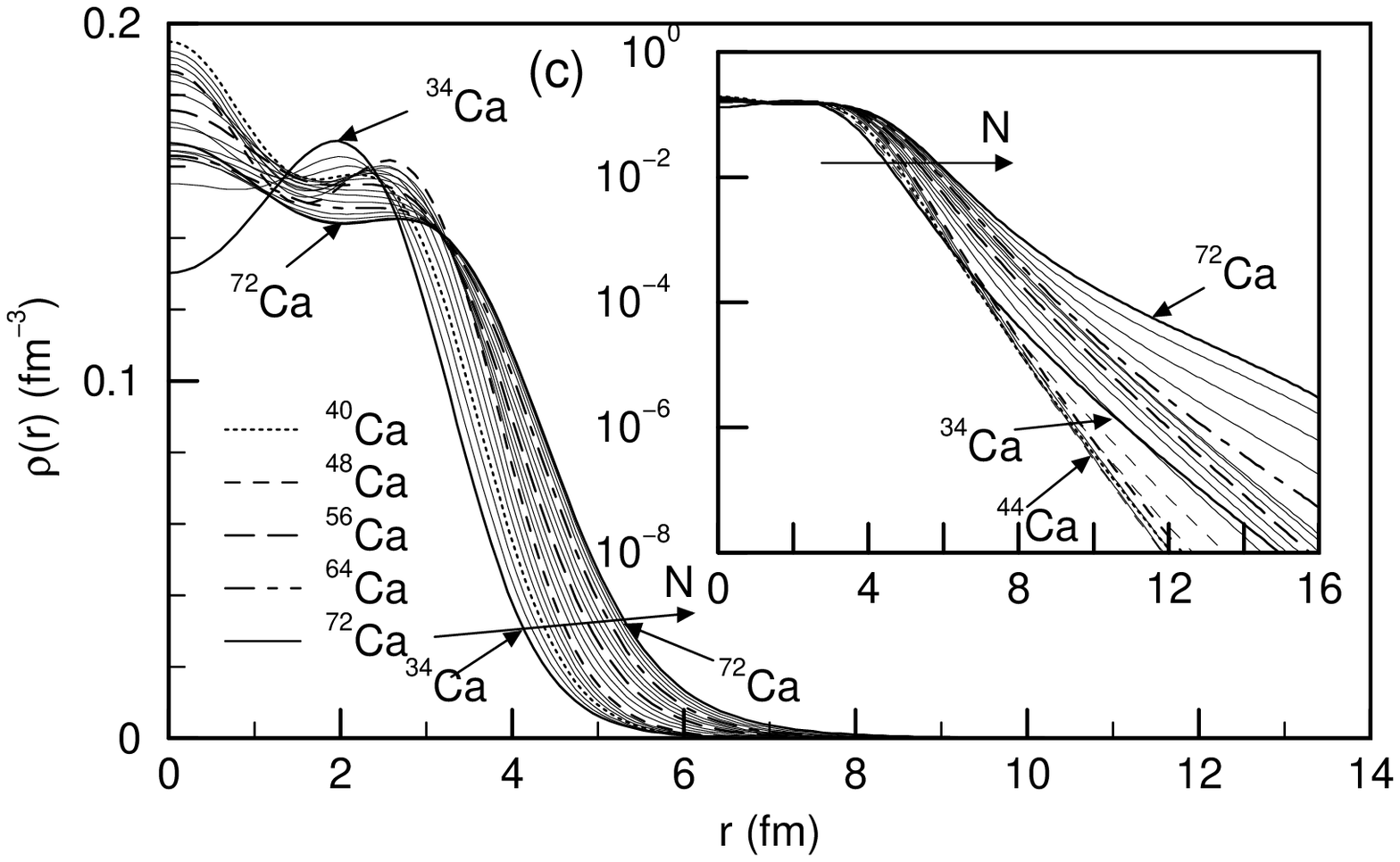,width=7.cm}}

\caption{(a) The neutron, (b) proton, and (c) matter density
distributions for the Ca isotopes. The arrows represent the
density variation tendency with the neutron number $N$.}
\label{fig. rho}
\end{figure}

The density distributions for some Ca isotopes are given for the
neutron, proton and matter densities in Fig. \ref{fig. rho}a-c,
respectively. In these figures, the arrows represent the density
variation tendency with the mass number $A$. From Fig. \ref{fig.
rho}c, the matter densities extend outwards with $A$ at the
nuclear surface ($3<r<6$ fm), while the variation is small in the
center ($r<3$ fm). For the proton density in Fig. \ref{fig. rho}b,
the density for $r<3$ fm decreases dramatically, and at the
surface ($r\sim 4$fm) extends outwards monotonously with $A$,
although the proton number is fixed to 20. It is in accordance
with the increase of the proton radii $r_p$ and the charge radii
$r_c$ as shown in Fig. \ref{fig. radii}. The increase of the
radius is just due to the nuclear saturation property, as the
total mass number $A$ of the nuclei is increasing. The neutron
densities, which are represented in Fig. \ref{fig. rho}a, extend
greatly outwards with $A$ on the surface and neutron skin is
developing. In the center, most neutron densities also increase
with $A$, however the increase is not as dramatic as the decrease
of the proton density. As a result, the matter densities vary
(mainly decrease) slightly in the interior region.

The halo phenomena are always related with the wide extension of
the nuclear density distributions in space. In order to exhibit
clearly the large spacial extension, the density distributions are
also shown on a logarithmic scale in the corresponding figures of
Fig. \ref{fig. rho}a, b, c as the inset for neutron, proton and
matter density distribution. From the inset figures of Fig.
\ref{fig. rho}a, the proton density decreases once more with $A$
in the rather exterior region of $r>8$ fm, meanwhile heavier Ca
isotopes extend toward outside in the region of $4<r<6$ fm and the
proton density decreases dramatically with $A$ in the center $r<3$
fm. So there exist two switching regions for the proton density:
one lies in the nuclear surface ($r\sim 4$ fm) due to the nuclear
saturation property, and the other appears over $6-8$ fm region,
which are related with the neutron-deficiency.

From the neutron density represented in the inset of Fig.
\ref{fig. rho}a, it is clearly seen that the neutron density
distributions for Ca isotopes with $N>40$ extend much more widely
than those for lighter ones. The tail becomes larger and larger
with the neutron number $N$ increased from 42 to 52, which is
quite different from the cases in Ni and Sn isotopes
\cite{ME98,MT99}. In the latter two isotope chains, the tails of
the neutron densities reach saturation beyond a specific nucleus,
i.e., $A=90$ for Ni and $A=160$ for Sn. However, no saturation
point is seen for the Ca isotopes as shown in Fig. \ref{fig.
rho}a.

For the tails of the matter density distributions in the inset of
Fig. \ref{fig. rho}c, they are mainly determined by the proton
densities for proton-rich nuclei while by the neutron densities
for neutron-rich nuclei. The nucleus with the smallest tail for
matter distribution is shown to be $^{44}$Ca, which is nearest to
the $\beta$-stability line, while not the lightest
neutron-deficiency nucleus $^{34}$Ca. For neutron rich nuclei with
$N$ from 42 to 52, the tail behaviors of $r_m$ are consistent with
that of $r_n$, which are strong evidences that giant neutron halos
exist in exotic Ca nuclei ($A>60$).

\begin{figure}
\centerline{\epsfig{figure=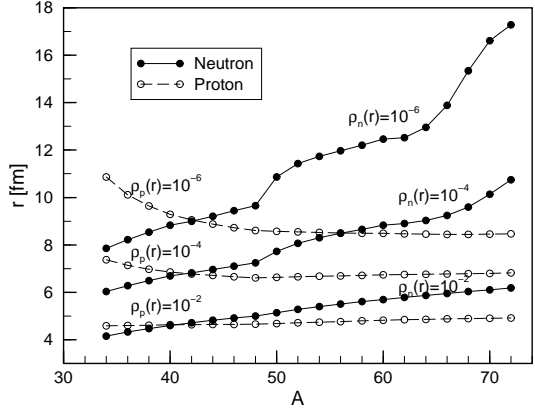,width=7.cm}}
 \caption{ The radii $r$ corresponding to different given proton and
neutron densities $\rho_p(r) ( \rho_n(r) ) = 10^{-2}, 10^{-4},$
and $10^{-6}$ fm$^{-3}$),respectively, as a function with the mass
number in Ca isotope chain.}
 \label{fig. r_rho}
\end{figure}

To give a simple quantitative idea on the density distributions
and halos, in Fig. \ref{fig. r_rho}, the radii at which the proton
and neutron densities, $\rho_p(r)$ and $\rho_n(r) =10^{-2},
10^{-4},10^{-6}~{\rm fm}^{-3}$, respectively, as a function of the
neutron number for the Ca isotope chain are shown. As the central
density of nuclear matter is about $\rho_0\sim$0.16 fm$^{-3}$ and
the proton and the neutron density is about a half of $\rho_0$,
the value $r$ for $\rho_{p(n)}=10^{-2}$ fm$^{-3}$ corresponds to
the densities decrease to 10$\%$ of the central density and the
surface of the nuclei. The value $r$ for $\rho_{p(n)}=10^{-6}$
fm$^{-3}$ corresponds to the tails of density distribution. From
this figure, for $^{42}$Ca, we see that the proton and neutron
densities have the same values at the same $r$, i.e., the neutron
density distribution is very similar to that of proton for the
$\beta$ stable nuclei. For neutron rich nuclei, if $\rho_{n} =
\rho_{p} $, the neutron radius $r$ is much bigger. For neutron
deficient nuclei, the opposite are seen. For $\rho_{n}=10^{-6}$
fm$^{-3}$, a increasing tendency for $r$ curve can be clearly
seen, particularly, the curve shows a strong kink at $N=62$. The
kink is also a important signal for the giant neutron halos, as
well as the $S_{2n}$ and neutron radii $r_n$.

\subsection{Single particle levels in canonical basis and
contribution of continuum}

The above results can be understood more clearly by considering
the microscopic structure of the underlying wave functions and the
single-particle energies in the canonical basis\cite{ME98}.  In
Fig. \ref{fig. n_level}, the neutron single-particle levels in the
canonical basis are shown for the even Ca isotopes from the mass
number $A=34$ to $A=72$. The shell closure ($N=20$ and $28$) and
sub-shell closure ($N=40$) are clearly seen as big gaps between
levels. A dotted line in the figure represents the neutron
chemical potential $\lambda_n$, which jumps three times at the
magic or submagic neutron number on its way to almost zero at
$^{72}$Ca. These jumps correspond to the shell (or sub-shell)
closure just as the appearance in the $S_{2n}$ case. The
$\lambda_n$ comes close to zero for nuclei near the neutron drip
line, i.e., $\lambda_n \approx -0.92$ MeV for $^{62}$Ca, -0.64 MeV
for $^{64}$Ca, -0.44 MeV for $^{66}$Ca, -0.28 MeV for $^{68}$Ca,
-0.16 MeV for $^{70}$Ca, and -0.07 MeV for $^{72}$Ca. Meanwhile,
no jump at $N=50$ in the $\lambda_n$ curve is in agreement with
the disappearance of this traditional magic number mentioned above
in the $S_{2n}$ case. From Fig. \ref{fig. n_level}, the big gap
between the $1g_{9/2}$ orbit and the $s-d$ shell has disappeared
in the neutron drip line region due to the lowering of $3s_{1/2}$
and $2d_{5/2}$ orbits. For example, this gap for $^{72}$Ca is only
1.02 MeV.

\begin{figure}
\centerline{\epsfig{figure=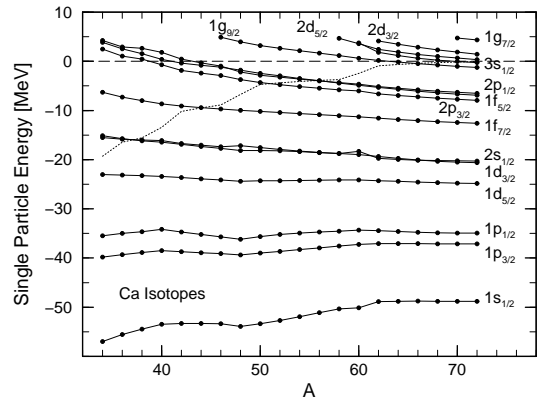,width=7.cm}} \caption{The
single-particle neutron energies in the canonical basis with the
mass number A for the Ca isotopes. The Fermi surfaces are shown by
the dot-dashed line.} \label{fig. n_level}
\end{figure}

As neutron chemical potential $\lambda_n$ approaches zero, the Ca
isotopes with $A>60$ are all weakly bound nuclei. This means that
the additional neutrons will occupy the weakly bound states, which
are very close to the continuum region. These neutrons supply very
small binding energies, and result in nearly vanishing two neutron
separation energies $S_{2n}$. Furthermore, the pairing interaction
scatters the neutron pairs from such weakly bound states to
continua as the Fermi level is close to zero. The single particle
occupation of continua then affects the nuclear properties. For Ca
isotopes with $A>60$, the added neutrons will occupy the weakly
bound states and continua: $1g_{9/2}, 3s_{1/2}, 2d_{5/2},
2d_{3/2}$, etc. Such orbits play importance roles as discussed
below.

\begin{figure}
\centerline{\epsfig{figure=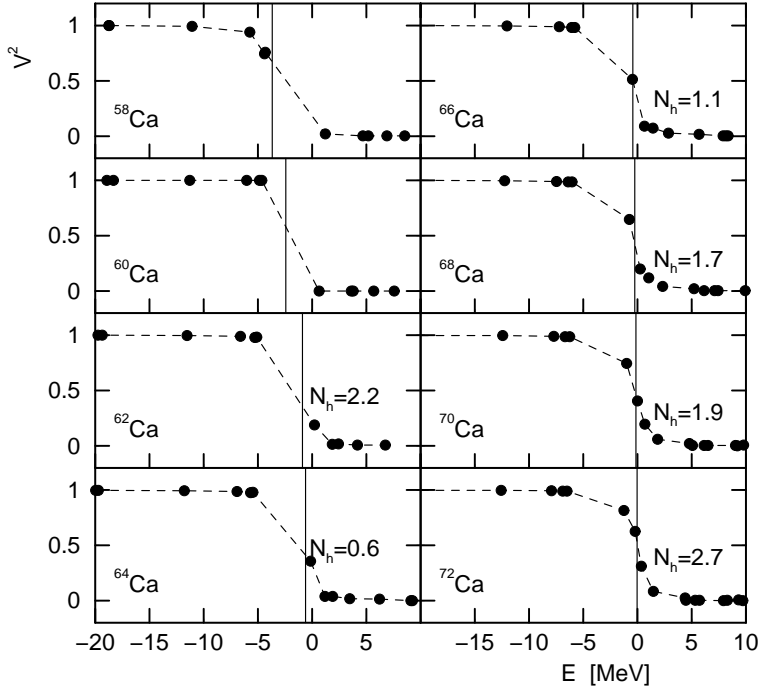,width=10.cm}} \caption{The
occupation probabilities in the canonical basis for various even
Ca isotopes as a function of the single particle energy. The
chemical potential is indicated with a vertical line. The number
$N_h$ of neutron in the halo are also shown.} \label{fig. occup}
\end{figure}

In Fig. \ref{fig. occup}, we present the occupation probabilities
$v^2$ of neutron levels near Fermi surface (i.e., $-20\leq E\leq
10$ MeV) in the canonical basis for several neutron-rich even Ca
isotopes, i.e., $^{58-72}$Ca. The neutron chemical potential
$\lambda_n$ is indicated by a vertical line. For nuclei with mass
number $A<60$, the chemical potential is quite large (e.g., -13.1
MeV for $^{40}$Ca, -6.74 MeV for $^{48}$Ca, and -3.69 MeV for
$^{58}$Ca), the occupation probabilities for the continua are
nearly zero. As the neutron number goes beyond the subshell
$N=40$, the Fermi surface approaches to zero and the occupation of
the continuum becomes more and more important. Summing up the
occupation probabilities $v^2$ for the states with $E>0$, one can
get the contribution of the continua $n_h$ \cite{MR98}. They are
approximately 2.2, 0.6, 1.1, 1.7, 1.9 and 2.7 for $^{62}$Ca,
$^{64}$Ca, $^{66}$Ca, $^{68}$Ca, $^{70}$Ca, and $^{72}$Ca,
respectively. In ref. \cite{MR98}, from two to roughly six
neutrons in total scattering in the continuum is predicted in Zr
isotopes with $N>82$, while the neutron number scattering to
continuum in Ca isotopes is reduced to about $1\sim 3$.

\begin{figure}
\centerline{\epsfig{figure=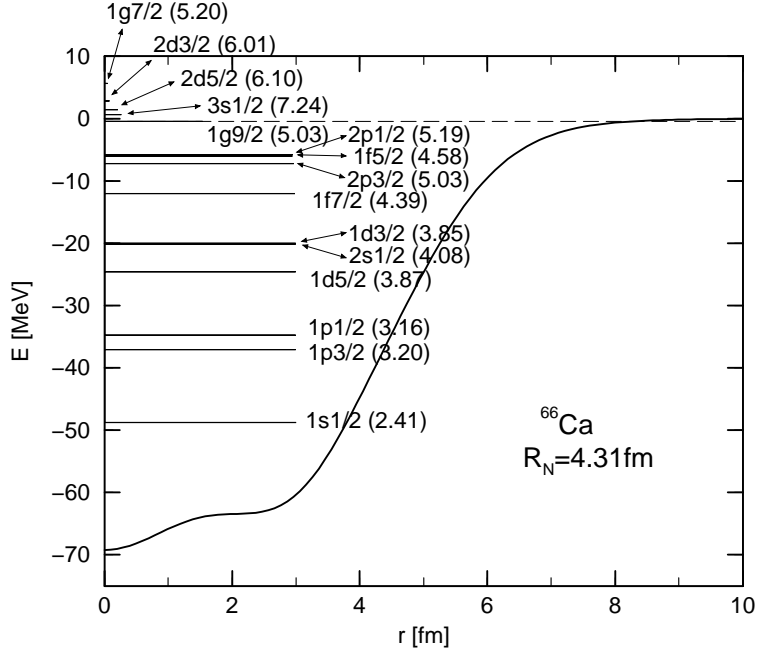,width=10.cm}} \caption{The
single-particle levels in the canonical basis for the neutron in
$^{66}$Ca. The neutron potentials $V(r)+S(r)$ is denoted by the
solid line and the Fermi surface is shown by a dashed line. The
occupied probabilities for the single-particle levels are shown by
their line-length. Also the root mean square radius $r_{nlj}$ in
fm for every level is listed in the parenthesis behind the
corresponding level.} \label{fig. ca66_lv}
\end{figure}

As a typical example, we investigate in detail the single particle
levels and contributions of continua with exotic nuclei $^{66}$Ca.
Fig. \ref{fig. ca66_lv} shows the neutron single-particle levels
in the canonical basis for $^{66}$Ca. The length of each level is
proportional to its occupied probability $v^2$. The Fermi surface
for neutron ($\lambda_n\approx -0.435$ MeV) is represented with a
dashed line, and the nuclear mean field potential $V(r)+S(r)$ is
denoted with the solid curve. We here define the root mean square
radius $r_{nlj}$ for one single-particle level denoted by $nlj$
as:
\begin{equation}
   r_{nlj}=\frac{\int \rho_{nlj} r^2d\tau}{\int \rho_{nlj}d\tau } ,
\end{equation}
where $\rho_{nlj}$ presents the probability density of each level
and the corresponding wave function $\psi_{nlj}$. This root mean
square radius $r_{nlj}$ for each level is shown in the parenthesis
behind the corresponding level label in units of fm. Also in Tab.
V, we list the single particle energies $\epsilon_{nlj}$, the rms
radii $r_{nlj}$ and the occupied probabilities $v^2_{nlj}$ for all
the neutron levels with energies $\epsilon < 10$ MeV in $^{66}$Ca
calculated from RCHB theory.

Here, the  state $1g_{9/2}$ is weakly bound with the energy -0.48
MeV, while the $3s_{1/2}$, $2d_{5/2}$, $2d_{3/2}$, and $1g_{7/2}$
levels are in the continuum with energies 0.64, 1.41, 2.85, and
5.67 MeV, respectively. We note that due to the absence of a
centrifugal barrier, the orbit $3s_{1/2}$ lies below states
$2d_{3/2}$ and $2d_{5/2}$.  The occupied probabilities $v^2$ of
these states are: 0.514 for $1g_{9/2}$, 0.089 for $3s_{1/2}$,
0.071 for $2d_{5/2}$, 0.027 for $2d_{3/2}$, and 0.017 for
$1g_{7/2}$. We get about 0.85 neutrons in these continuum states,
which are about $4/5$ of the total contribution of the continuum
$n_h=1.1$ for $^{66}$Ca (see Fig. \ref{fig. occup}). The
$3s_{1/2}$ state has a rms radius 7.24 fm in comparison with the
rms radii of the neighbor states ($\sim5$ fm) and the total
neutron rms radii (4.314 fm) owing to its zero centrifugal
barrier. Therefore the nucleon occupying on the state $3s_{1/2}$
will contribute to the nuclear rms radius considerably.

The relative contributions $\rho_{nlj}/\rho_n$ of different
single-particle orbits to the full neutron density as a function
of radical distance $r$ are represented in Fig. \ref{Fig.
rho_rhon} for the typical nuclei $^{66}$Ca. For comparison we also
present the total neutron density $\rho_n$ with the shaded area in
arbitrary units. In the interior of nuclei ($r< 4$ fm), the
contribution to neutron distribution mainly comes from the
low-lying energy levels such as $1s_{1/2},~1p_{3/2},~1p_{1/2},
~1d_{5/2}, ~2s_{1/2}, ~1d_{3/2}$. In the nuclear surface($r\sim 5$
fm) and the beginner part of the tail ($6<r<10$ fm), the weakly
bound $f$-$p$ shell states and very weakly $1g_{9/2}$ orbit play
dominant role for neutron distribution. Then their contributions
gradually decrease with the radius, while the contributions from
the continuum (e.g. $3s_{1/2}, 2d_{5/2}$) become more and more
important. For $r>15$ fm, the state $3s_{1/2}$ is dominant (about
60 percent) owing to its zero centrifugal barrier, and the other
continuum states $2d_{5/2}, 2d_{3/2}, 4s_{1/2}$ and the very
weakly bound state $1g_{9/2}$ also play their roles. It is well
shown that the contributions of the continua are crucial to the
tail, which is closely connected with the halo phenomena.

\begin{table}[htbp]
\begin{center}
\caption{The properties of neutron single-particle levels in
$^{66}$Ca, including single particle energies, occupation
probabilities, and rms radii for the single-particle levels.}
\begin{tabular}{c|ccc||c|cccc} \hline\hline
$nlj$ & $\epsilon$ & $v^2$ & $r$ & $nlj$ & $\epsilon$ & $v^2$ &
$r$ & \\ \hline
$1s_{1/2}$ & -48.76 & 1.00 & 2.41 & $3s_{1/2}$ &  0.64  & 0.089 & 7.24 & \\
$1p_{3/2}$ & -37.08 & 1.00 & 3.20 & $2d_{5/2}$ &  1.41  & 0.071 & 6.10 & \\
$1p_{1/2}$ & -34.74 & 1.00 & 3.16 & $2d_{3/2}$ &  2.85  & 0.027 & 6.01 & \\
$1d_{5/2}$ & -24.56 & 1.00 & 3.87 & $1g_{7/2}$ &  5.67  & 0.017 & 5.20 & \\
$2s_{1/2}$ & -20.11 & 1.00 & 4.08 & $4s_{1/2}$ &  7.88  & 0.002 & 8.42 & \\
$1d_{3/2}$ & -20.04 & 1.00 & 3.85 & $3p_{3/2}$ &  8.11  & 0.002 & 6.51 & \\
$1f_{7/2}$ & -12.05 & 1.00 & 4.39 & $3p_{1/2}$ &  8.35  & 0.002 & 6.63 & \\
$2p_{3/2}$ & -7.24  & 1.00 & 5.03 & & & & & \\
$1f_{5/2}$ & -6.07  & 1.00 & 4.58 & & & & & \\
$2p_{1/2}$ & -5.79  & 1.00 & 5.19 & & & & & \\
$1g_{9/2}$ & -0.48  & 0.514 & 5.03 & & & & & \\ \hline\hline
\end{tabular}
\end{center}
\end{table}
In Tab. VI, the energies, rms radii and  occupied probabilities of
the neutron single-particle level $3s_{1/2}$ for even isotopes
$^{60-72}$Ca isotopes are presented. Its energy decreases with
neutron number $N$, i.e., it lies in continuum for $^{60-68}$Ca
and becomes slightly bound for $^{70, 72}$Ca. As both the rms
radii and $v^2$ for the state $3s_{1/2}$ increase monotonously
with the mass number $A$, it will drive the tail of neutron
density distribution become larger and larger as shown above. Of
course, with more neutrons added to $^{66}$Ca up to the heaviest
bound nucleus predicted as $^{72}$Ca, the contributions to the
tail from the other continuum $2d_{5/2}, 4s_{1/2}$ and $2d_{3/2}$
will also become more important due to their larger occupation.

\begin{figure}
\centerline{\epsfig{figure=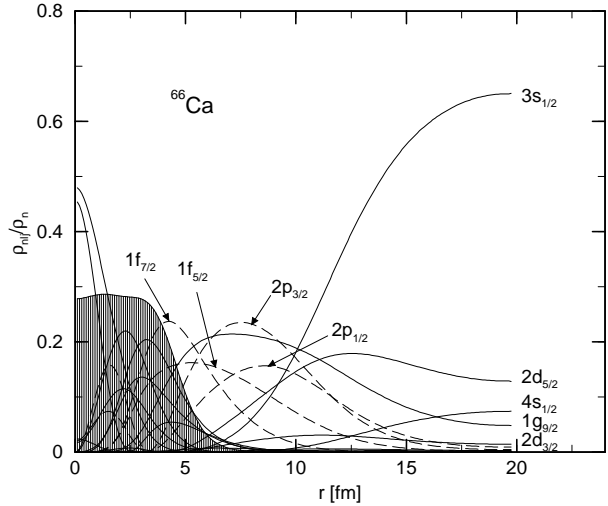,width=8.cm}} \caption{
Relative contributions of the different orbits to the total
neutron density as a function of the radius. The shaded area
indicates the total neutron density in arbitrary units.}
\label{Fig. rho_rhon}
\end{figure}

\begin{table}[htbp]
\begin{center}
\caption{The properties of $3s_{1/2}$ for $^{60-72}$Ca}
\begin{tabular}{cccccccc}
\hline \hline A & 60 & 62 & 64 & 66 & 68 & 70 & 72 \\ \hline
$\epsilon$~(MeV)  & 3.751 & 1.842 & 1.162 & 0.638 & 0.247 & -0.028 & -0.238 \\
$r$~(fm)  & 5.524 & 6.356 & 6.781 & 7.240 & 7.709 & 8.062  & 8.212 \\
$v^2$    & $\sim$0.0   & 0.015 & 0.039 & 0.089 & 0.201 & 0.404  & 0.628 \\
\hline\hline
\end{tabular}
\end{center}
\end{table}

\subsection{Vector and scalar potentials and their isospin evolution}

One of the essential differences between a non-relativistic single
particle equation and the relativistic Dirac equation in nuclear
physics is the fact that the relativistic equation contains from
the beginning two potentials $V$ and $S$, which have different
behaviors under Lorentz transformation \cite{RI96}. We now
investigate the vector and scalar potentials in the Ca isotope
chains in detail. Fig. \ref{Fig.VS} presents the vector $V$ and
scalar potential $S$ for proton (left panels) and neutron (right
panels) in Ca chains as a function of the radial distance $r$,
respectively. We can see from the figure that both the vector and
scalar potentials have a Woods-Saxon shape, i.e., the similar
nuclear density distribution. They nearly vanish outside the
nucleus and they are more or less constant in the nuclear
interior, namely $S\approx -420 $ MeV as an attractive potential
and $V \approx 350$ MeV as an repulsive potential.

\begin{figure}
\centerline{\epsfig{figure=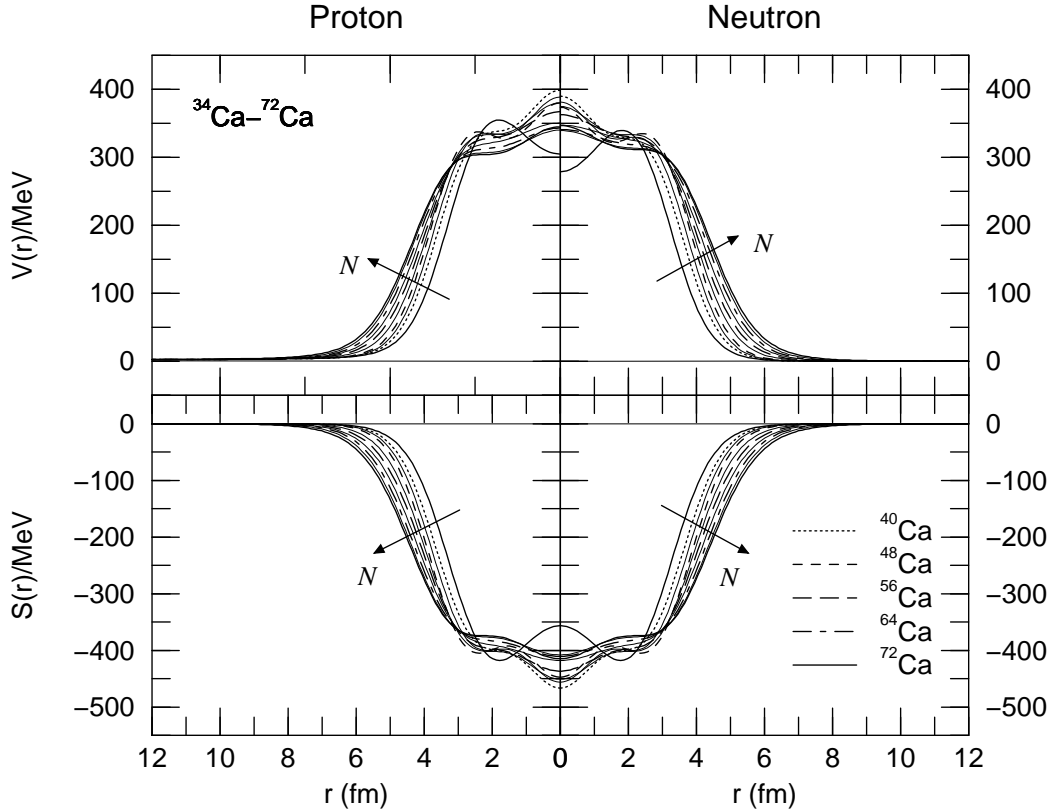, width=14.cm}} \caption{The
vector and scalar potentials (V(r),S(r)) for the Ca isotopes. The
left panel presents those for proton, while the right panel for
neutron.} \label{Fig.VS}
\end{figure}

We can clearly see that the scalar potential $S(r)$ for proton is
just the same as that for neutron in all cases. It is isospin
independent as the scalar potential comes from $\sigma$ meson
field, which takes the same form for either proton or neutron. On
the other hand, the vector potential $V(r)$ is isospin dependent:
the potential for proton is somewhat different from neutron. The
difference comes from the $\rho$ meson and the Coulomb
interaction. For a given nucleus, the vector potential for proton
is a little larger than that for neutron in the nuclear interior,
and it does not vanish outside the nucleus due to the long range
Coulomb interaction.

\subsection{Spin-orbit splitting in exotic nuclei}

An advantage of the relativistic mean field theory is to get the
spin-orbit coupling naturally. The spin-orbit splitting has been
discussed through various themes with relativistic or
non-relativistic microscopic theory \cite{MT99,LO00,LA98}. Now we
examine the spin-orbit splitting for the whole even Ca isotopes in
the relativistic continuum Hatree-Bogoliubov theory.

The spin-orbit splitting energy $E_{ls}$ for the two partners
($nlj=l-1/2, nlj=l+1/2$) is defined as:
\begin{equation}
E_{ls}=\frac{E_{nlj=l-1/2}-E_{nlj=l+1/2}}{2l+1}.
\end{equation}
In Fig. \ref{Fig. CaSO}, the spin-orbit splitting energies
$E_{ls}$ in Ca isotopes are shown as a function of mass number $A$
for the proton (lower panel) and neutron (upper panel) spin-orbit
partners ($1d_{3/2}, 1d_{5/2}$), ($1g_{7/2}, 1g_{9/2}$),
($1p_{1/2}, 1p_{3/2}$), ($1f_{5/2}, 1f_{7/2}$) and ($2p_{1/2},
2p_{3/2}$). For the lighter Ca isotopes, some doublets sit in the
continuum region with large positive energies in the canonical
basis. It would result in some uncertainty in spin-orbit
splitting. Therefore we limit the level energies $E$ below 10 MeV.
In Fig. \ref{Fig. CaSO}, the splitting decreases monotonically
from proton-rich side to the neutron rich-side for most doublets
except the doublets ($1p_{1/2}, 1p_{3/2}$) and ($2p_{1/2},
2p_{3/2}$). The splitting of the $p$ doublets between two
closures, i.e.,$^{40}$Ca, $^{48}$Ca, fluctuates quite a lot. Above
$N=28$, the splitting increases a little, then decline with the
neutron number. For the doublets ($1p_{1/2}, 1p_{3/2}$) and
($1d_{3/2}, 1d_{5/2}$), the spin-orbit splitting for neutron and
proton is very close to each other. For the doublets ($1f_{5/2},
1f_{7/2}$) and ($2p_{1/2}, 2p_{3/2}$), there are quite big
difference. Moreover, the splitting for neutron is usually smaller
than that for proton.

\begin{figure}
\centerline{\epsfig{figure=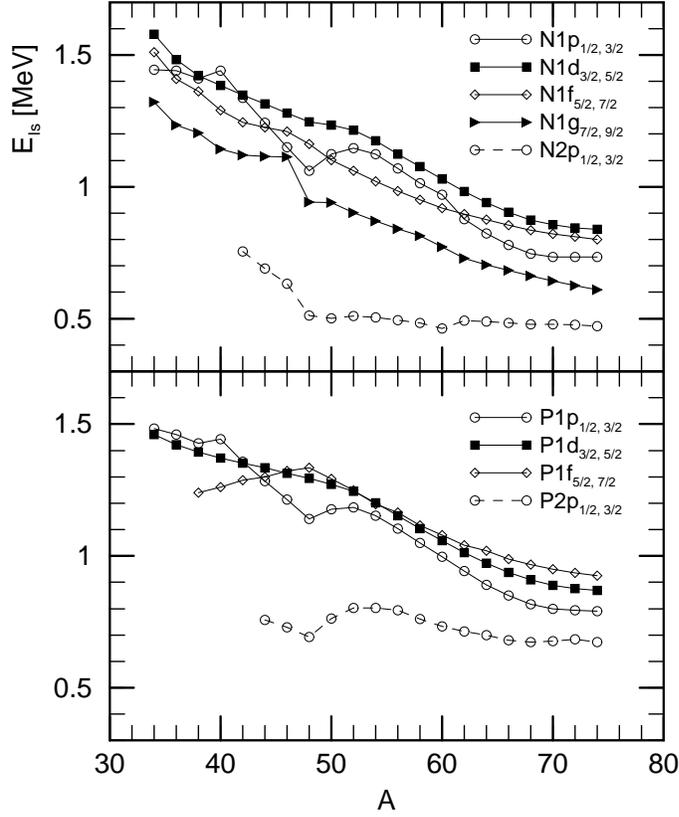,width=9.cm}} \caption{The
spin-orbit splitting energies $E_{ls}$ in Ca isotopes as a
function of mass number $A$ for the proton and neutron spin-orbit
doublets ($1d_{3/2}, 1d_{5/2}$), ($1g_{7/2}, 1g_{9/2}$),
($1p_{1/2}, 1p_{3/2}$), ($1f_{5/2}, 1f_{7/2}$) and ($2p_{1/2},
2p_{3/2}$).} \label{Fig. CaSO}
\end{figure}

It would be very helpful to examine the origin of the spin-orbit
splitting in the Dirac equation. For the Dirac nucleon moving in a
scalar and vector potentials, its equation of motion could be
decoupled and reduced to either the upper component or the lower
component, respectively. If it is reduced to the lower component,
it will be related with another interesting topic -- the
pseudo-spin symmetry \cite{MSY98}. Here we pay attention to the
spin-orbit splitting, which is related to the upper component. The
Dirac equation can be reduced for the upper component as follows,
\cite{MT99}:
\begin{eqnarray}
   & &  [ \frac {d^2} {dr^2}  - \frac 1  {E + 2M - V_V  + V_S}
        \frac {d(2M-V_V+V_S)} {dr} \frac {d} {dr} ] G^{lj}_i(r) \nonumber \\
   &-&  [  \frac { \kappa ( 1 + \kappa ) }  {r^2}
        - \frac 1 {E + 2M - V_V + V_S} \frac {\kappa} r
        \frac {d(2M - V_V + V_S)} {dr}  ] G^{lj}_i (r) \nonumber \\
 = &-&  (E + 2M - V_V + V_S ) ( E - V_V - V_S ) G^{lj}_i (r),
\label{larspinor4}
\end{eqnarray}
where
\begin{eqnarray}
       \kappa =
        \left\{ \begin{array}{cc}
            -l-1,  &  j=l+1/2  \\
            l,     &  j=l-1/2
         \end{array} \right.
\end{eqnarray}
The spin-orbit splitting is provided by the corresponding
spin-orbit term \cite{MT99}:
\begin{equation}
   \displaystyle \frac 1 {E + 2M- V(r) + S(r)} \frac {\kappa} r
        \frac {d(2M - V(r) + S(r))} {dr}
\label{spp}
\end{equation}
It can be seen clearly that the spin-orbit splitting is energy
dependent and depends also on the derivative of the potential $2M-
V(r) + S(r)$ as well as the particle distribution. Therefore the
so-called spin-orbit potential in the RCHB theory is defined
\cite{MT99}: $$V_{ls} = \frac{1}{r} \displaystyle \frac {d(2M -
V(r) + S(r))} {dr}$$.

The derivative of the potential difference,
$\dfrac{d(V(r)-S(r))}{dr}$, for proton and neutron in the Ca
isotopes is given in Fig. \ref{Fig. CadVmS}. The potential
difference, $V(r)-S(r)$, for both proton and neutron are almost
the same, as $V(r)-S(r)$ is a big quantity ( $\sim 700$ MeV ), and
the difference in the spin-orbit potential for proton and neutron
could be neglected. Therefore the proton and neutron $V_{ls}$ are
almost the same in the present model. That is the reason why the
spin-orbit splitting for neutron and proton doublets ($1p_{1/2},
1p_{3/2}$) and ($1d_{3/2}, 1d_{5/2}$) is very close to each other
as shown in Fig. \ref{Fig. CaSO}.

\begin{figure}
\centerline{\epsfig{figure=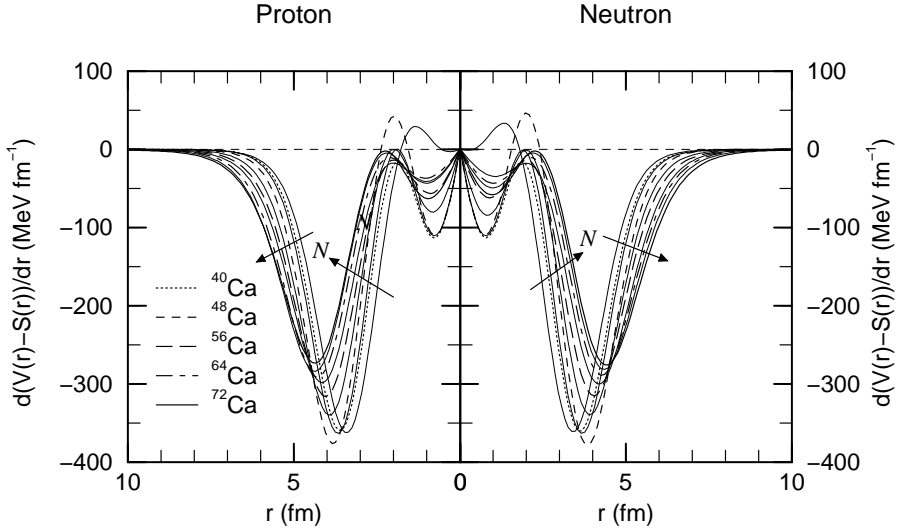,width=12.cm}} \caption{The
derivative of the potential difference,
$\dfrac{d(V(r)-S(r))}{dr}$, for proton (left) and neutron (right)
in the Ca isotopes as a function of the radius.} \label{Fig.
CadVmS}
\end{figure}


From $^{34}$Ca to $^{48}$Ca, the amplitude of $V_{ls}$ increases
monotonically and from $^{48}$Ca to $^{72}$Ca, the amplitude
decreases monotonically due to surface diffuseness. Meanwhile, the
maximum point of the potential $V_{ls}$ has an outwards tendency.
Thus the systematic decrease of spin-orbit splitting is partially
related with the decrease of $V_{ls}$. Furthermore, the systematic
decrease also comes from the diffuseness of the nuclear potential.
In Figs. \ref{Fig. CaVpS} and \ref{Fig. CaVmS}, we see that the
potentials $V(r)-S(r)$ and  $V(r)+S(r)$ extend outwards into the
nuclear surface, which makes the diffuseness increase with the
neutron number.

\begin{figure}
\centerline{\epsfig{figure=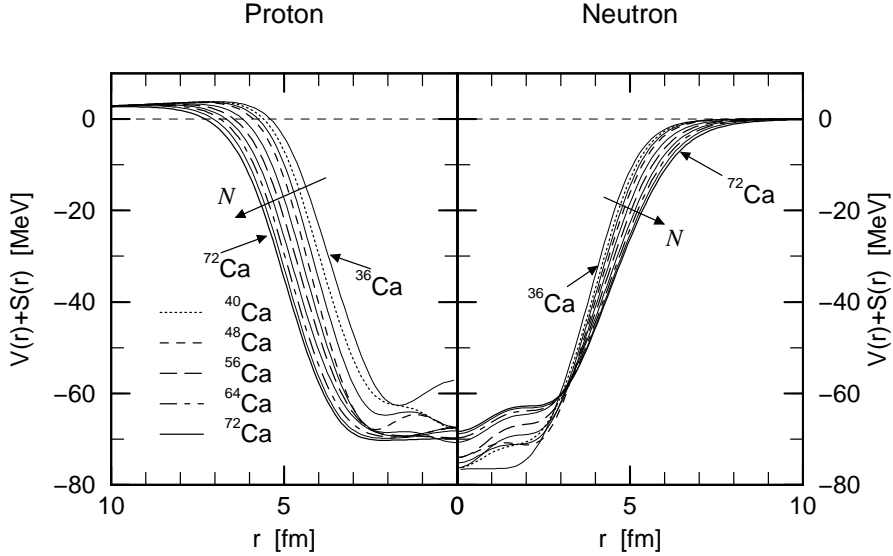,width=12.cm}} \caption{The
summed mean field potential, $V(r)+S(r)$, for proton and neutron
in even Ca isotopes. Directions of arrows in the figure show the
diffuseness of nuclear potential. } \label{Fig. CaVpS}
\end{figure}

\begin{figure}
\centerline{\epsfig{figure=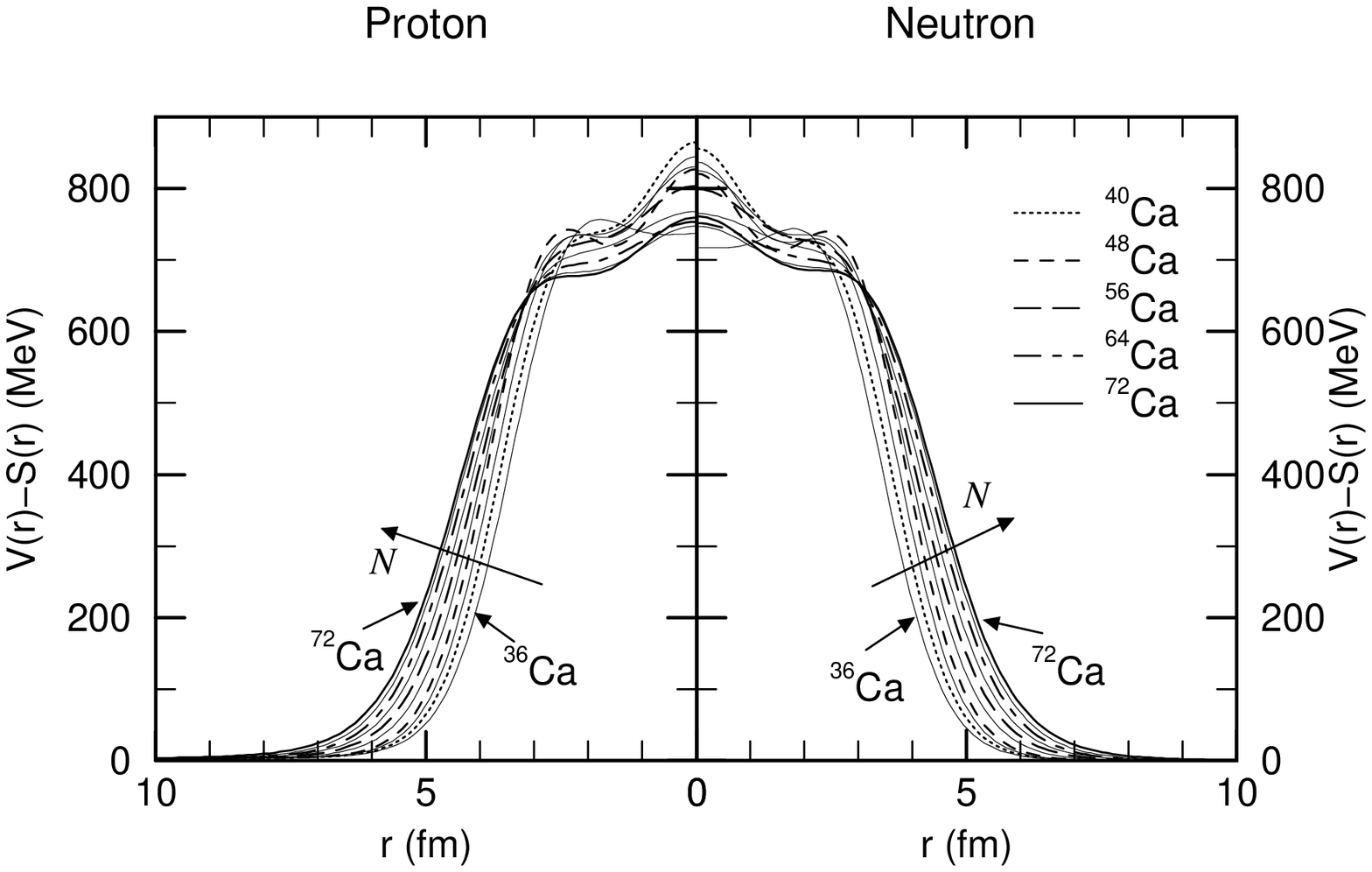,width=12.cm}} \caption{The
mean field anti-nucleon potential, $V(r)-S(r)$, for proton and
neutron in even Ca isotopes. Directions of arrows in the figure
show the diffuseness of nuclear potential.} \label{Fig. CaVmS}
\end{figure}

\begin{figure}
\begin{minipage}[t]{0.45\textwidth}
\centering
\includegraphics[width=0.8\textwidth]{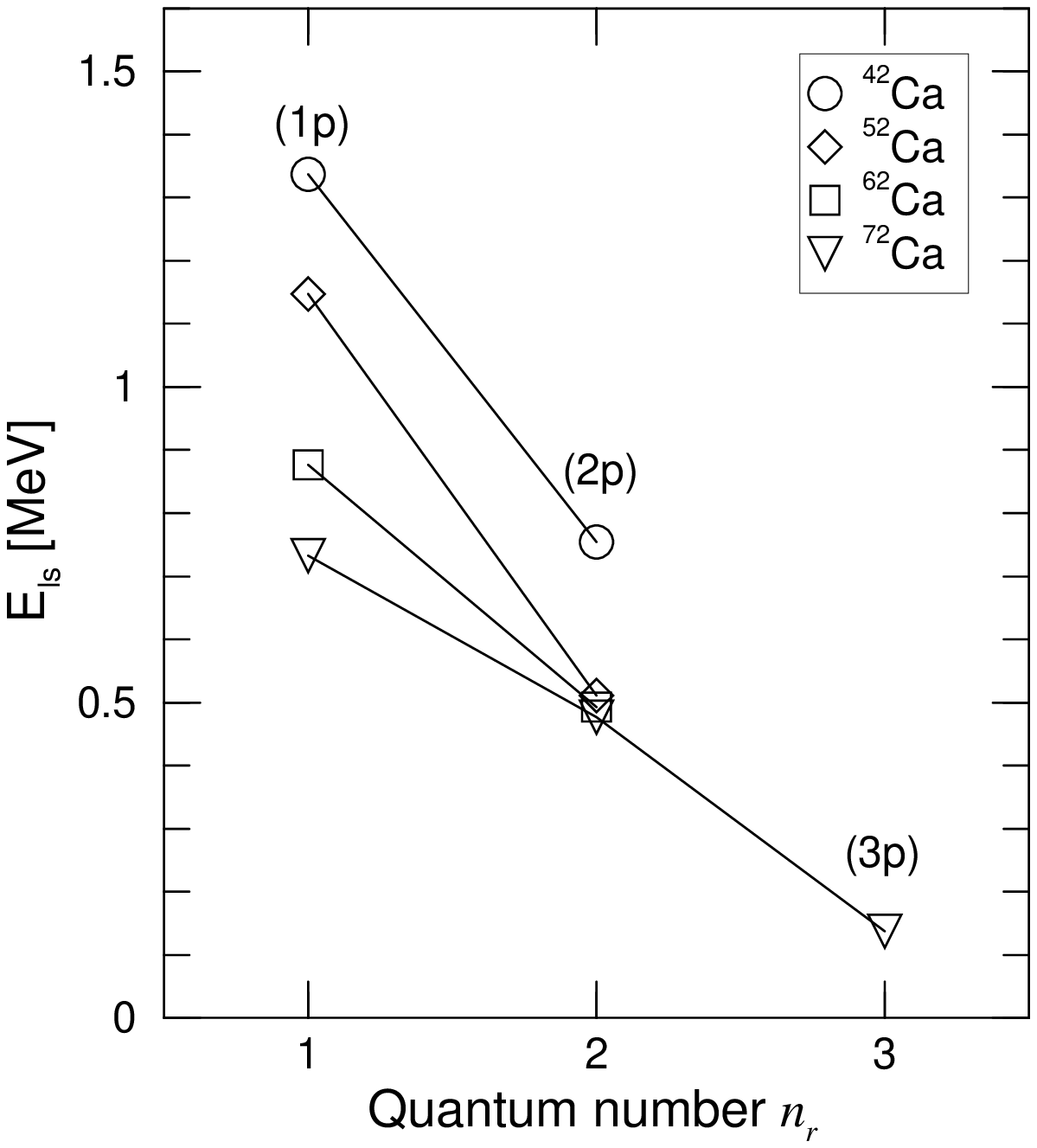}
\end{minipage}
\begin{minipage}[t]{0.45\textwidth}
\centering \includegraphics[width=0.8\textwidth]{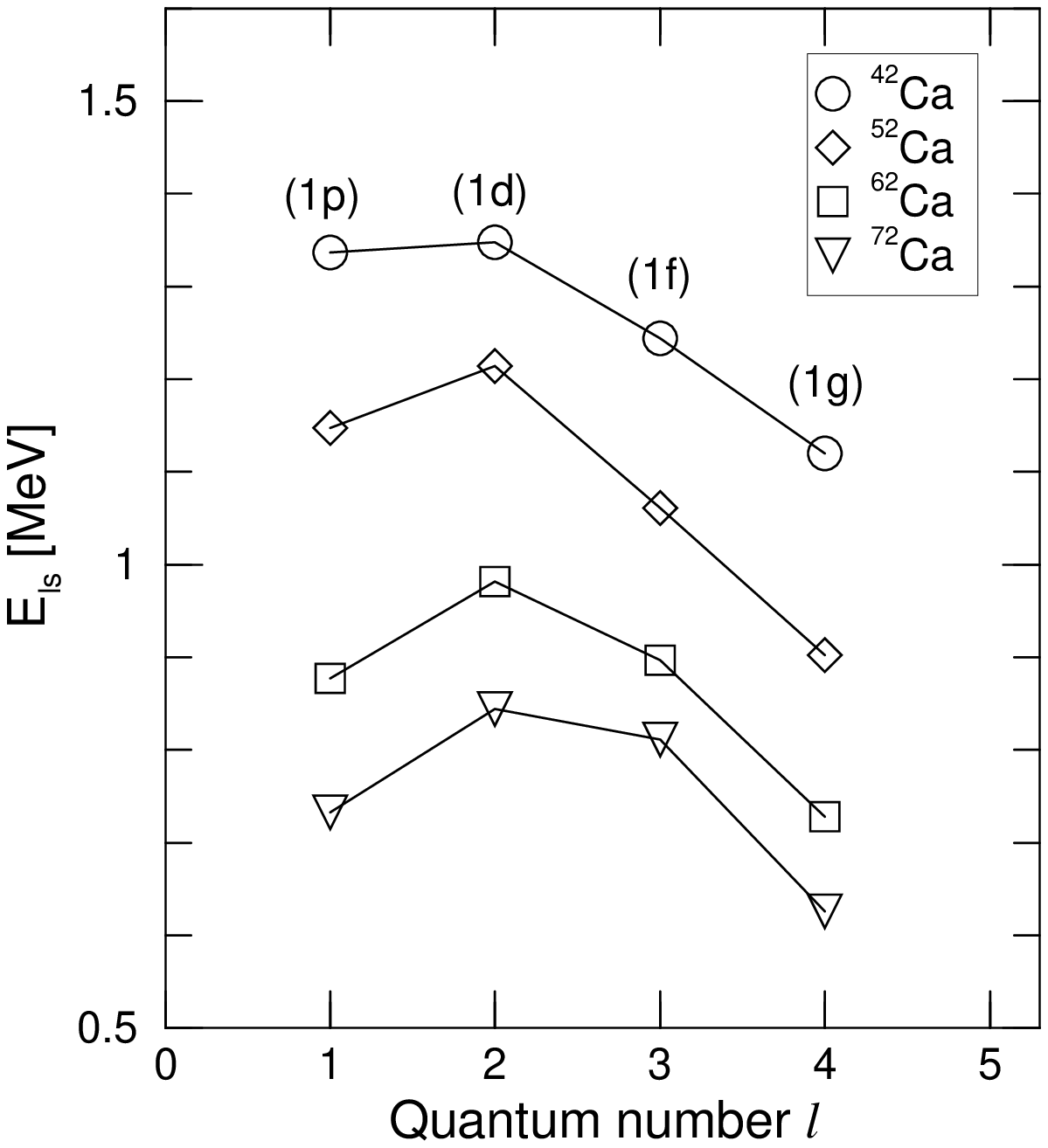}
\end{minipage}
 \caption{The spin-obit splitting $E_{ls}$ as a function of
a specific quantum number in $^{42,52,62,72}$Ca. The left panel
shows the splitting for $p$ orbits with the quantum number $n$;
the right panel shows the splitting for $n=1$ orbits with the
quantum number $l$.} \label{sosQn}
\end{figure}

So far we have investigated how the proton and neutron spin-orbit
splitting changes with the mass number $A$, i.e. the isospin
dependence. The decreasing tendency is mainly caused by the
derivative of anti-nucleon potential and the diffuseness of the
mean field. For one special nucleus, the potential is the same for
different spin-orbit doublets, so it is necessary to investigate
how the splitting changes with different quantum numbers, as in
Fig. \ref{sosQn}. To avoid the confusion caused by showing too
many doublets in the figure, we present the splitting energies for
$E_{ls}$ of $(p_{1/2}, p_{3/2})$ partners as a function of quantum
number $n$ in the left panel, and those of different $n=1$
doublets as a function of quantum number $l$ in the right panel,
respectively, for $^{42,52,62,72}$Ca.

In the left panel of Fig. \ref{sosQn}, the spin-orbit splitting in
doublets $(1p_{1/2}, 1p_{3/2})$ is much larger than that in
$(2p_{1/2}, 2p_{3/2})$ cases, and that in $(3p_{1/2}, 3p_{3/2})$
for $^{72}$Ca is the smallest. It can be clearly seen in
Eqn.(\ref{larspinor4}) that, besides the spin-orbit potential, the
spin-orbit term also depends on the factor,
$\dfrac{1}{E+2M-V(r)+S(r)}$. However, this factor has little
energy $E$ dependence. In fact, the energy E (varying from -40 to
10 MeV) is  a small quantity compared with the potential
$2M-V(r)+S(r)$, which is $2M$ ($\sim 1800$ MeV) in surface or
about 1000MeV in the nuclear interior. Thus the energy dependence
of the spin-orbit splitting difference caused directly by the
energy factor can be neglected.

It has been suggested that the  spin-orbit splitting difference
mainly comes from the overlap between the density distribution and
the  spin-orbit potential $V_{ls}$, as demonstrated for Sn
isotopes in Ref.\cite{MT99}. In Fig. \ref{soslapn}, for $^{42}$Ca,
$^{52}$Ca, $^{62}$Ca and $^{72}$Ca, the derivative of the
potential $V(r)-S(r)$, $d(V(r)-S(r))/dr$, and the density
distribution of the $p$-waves are given in the respective lower
panel, while the spin-orbit potential, $V_{ls}$, multiplied by the
density distributions for $p_{1/2}$ are respectively given in the
upper panel. The overlap between the spin-orbit potential $V_{ls}$
and the particle distribution is represented by the curve in the
corresponding upper panel.  Their contribution to the spin-orbit
splitting is proportional to the area surrounded by the
corresponding curve and the $x-$axis. It is clearly seen that the
overlap of $1p_{1/2}$ is larger than that of $2p_{1/2}$ for all
these isotopes and that of $3p_{1/2}$ in $^{72}$Ca is the
smallest. The overlap values of $2p_{1/2}$ for $^{52,62,72}$Ca are
very close from each other and all are smaller than that for
$^{42}$Ca. These observations can explain the features observed in
the right panel of Fig. \ref{sosQn}.

\begin{figure}[htbp]
\centering
\begin{minipage}[t]{0.45\textwidth}
\centering
\includegraphics[width=0.8\textwidth]{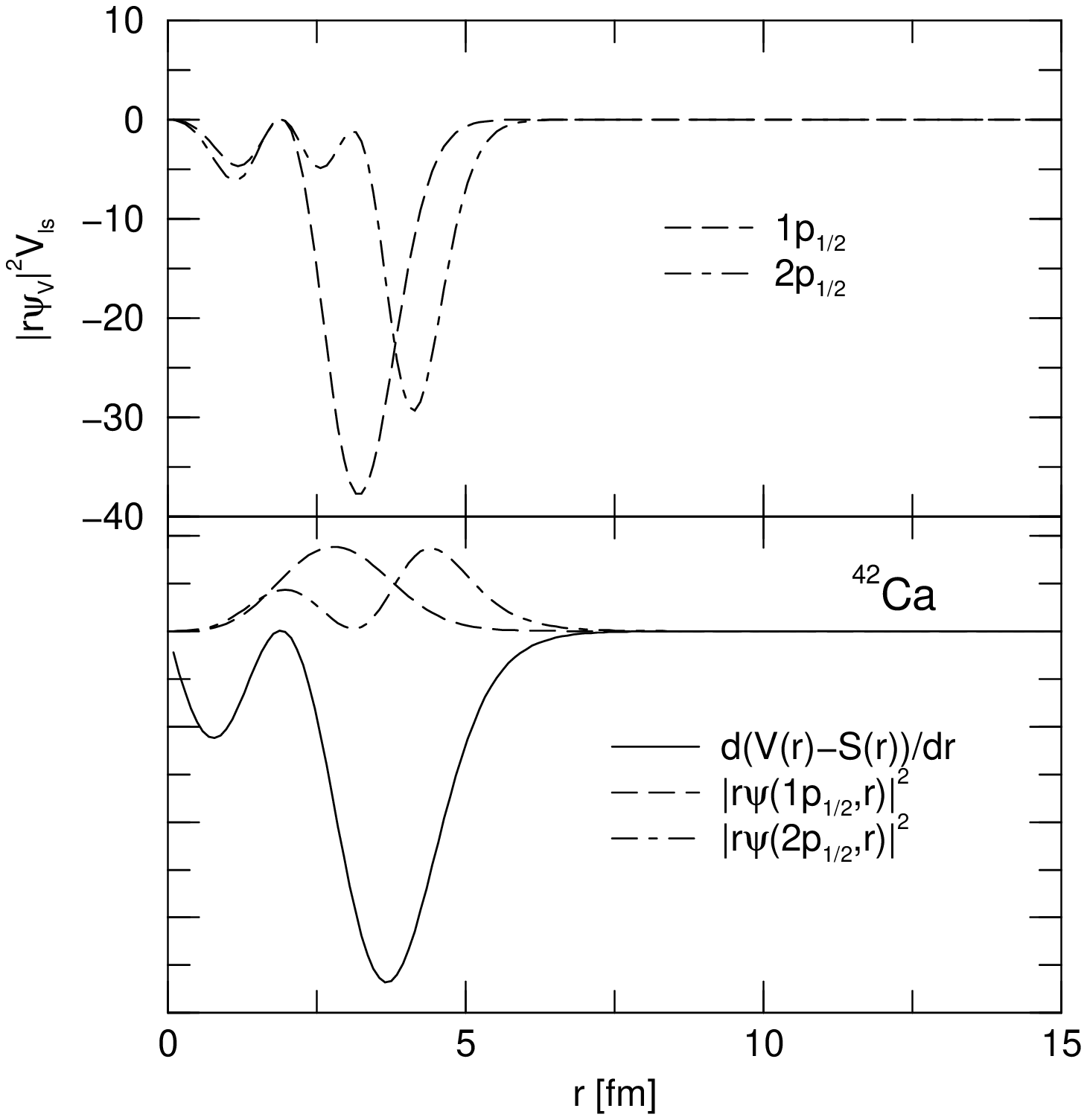}
\end{minipage}
\begin{minipage}[t]{0.45\textwidth}
\centering
\includegraphics[width=0.8\textwidth]{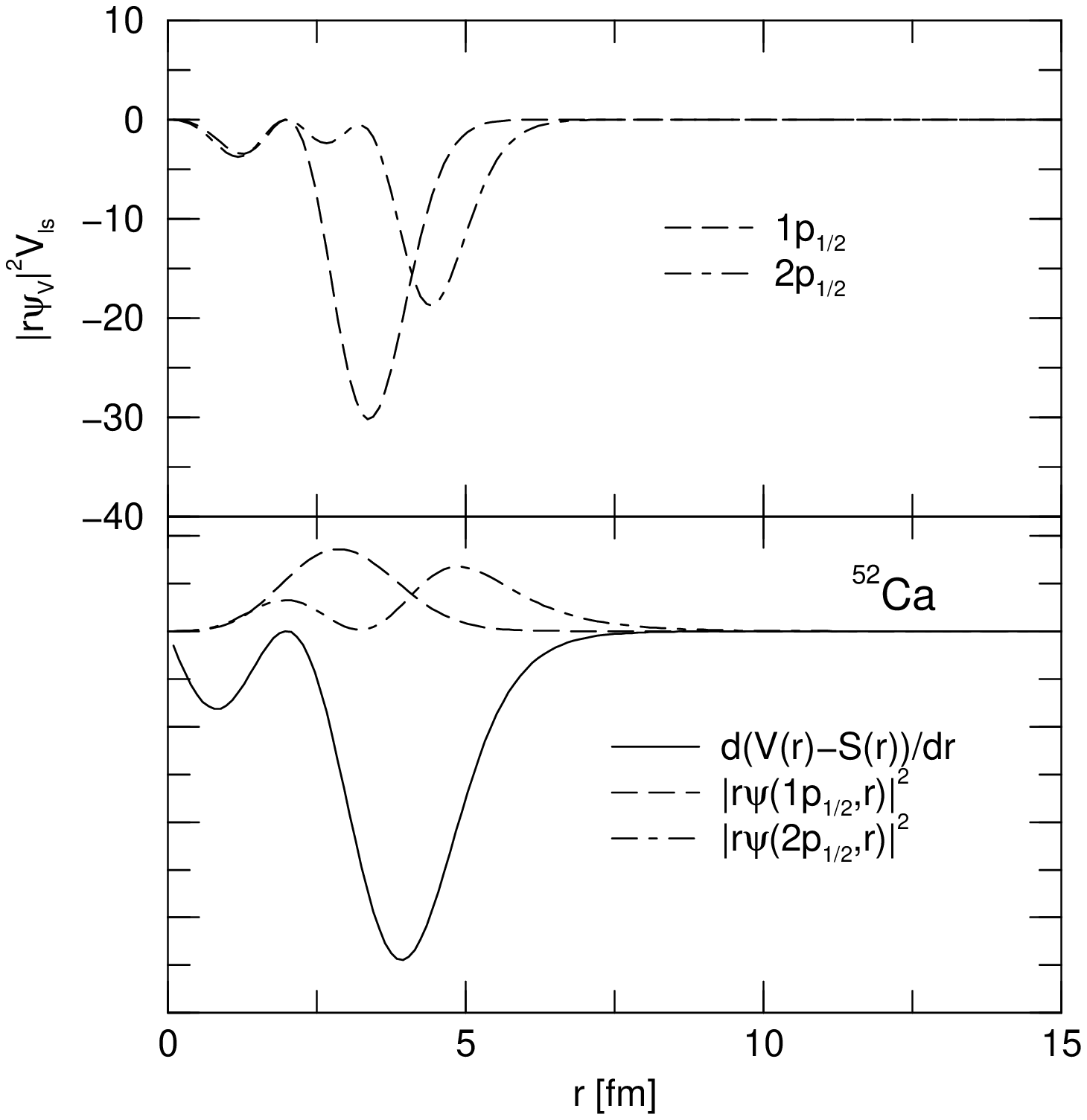}
\end{minipage}
\begin{minipage}[t]{0.45\textwidth}
\centering
\includegraphics[width=0.8\textwidth]{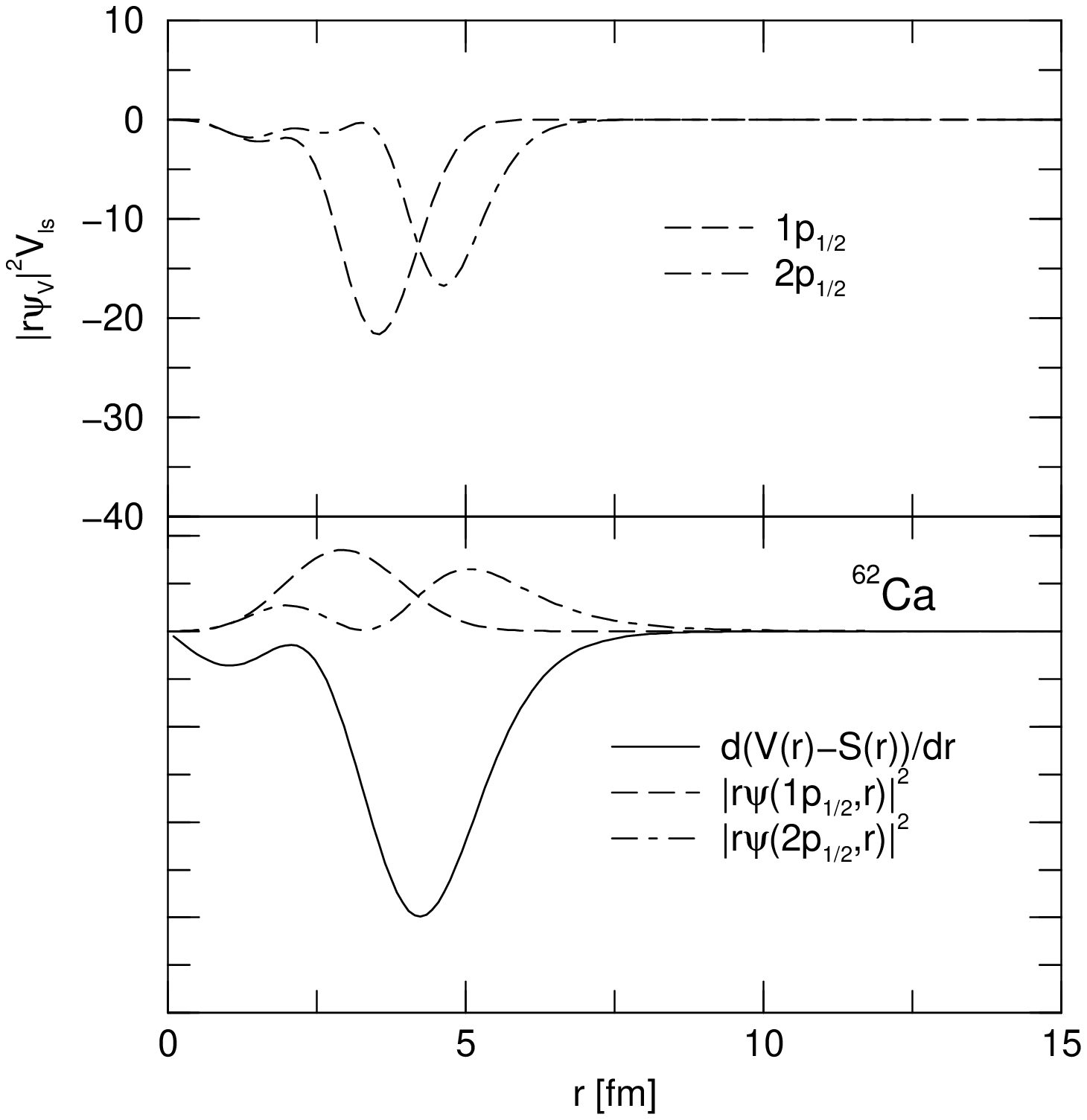}
\end{minipage}
\begin{minipage}[t]{0.45\textwidth}
\centering
\includegraphics[width=0.8\textwidth]{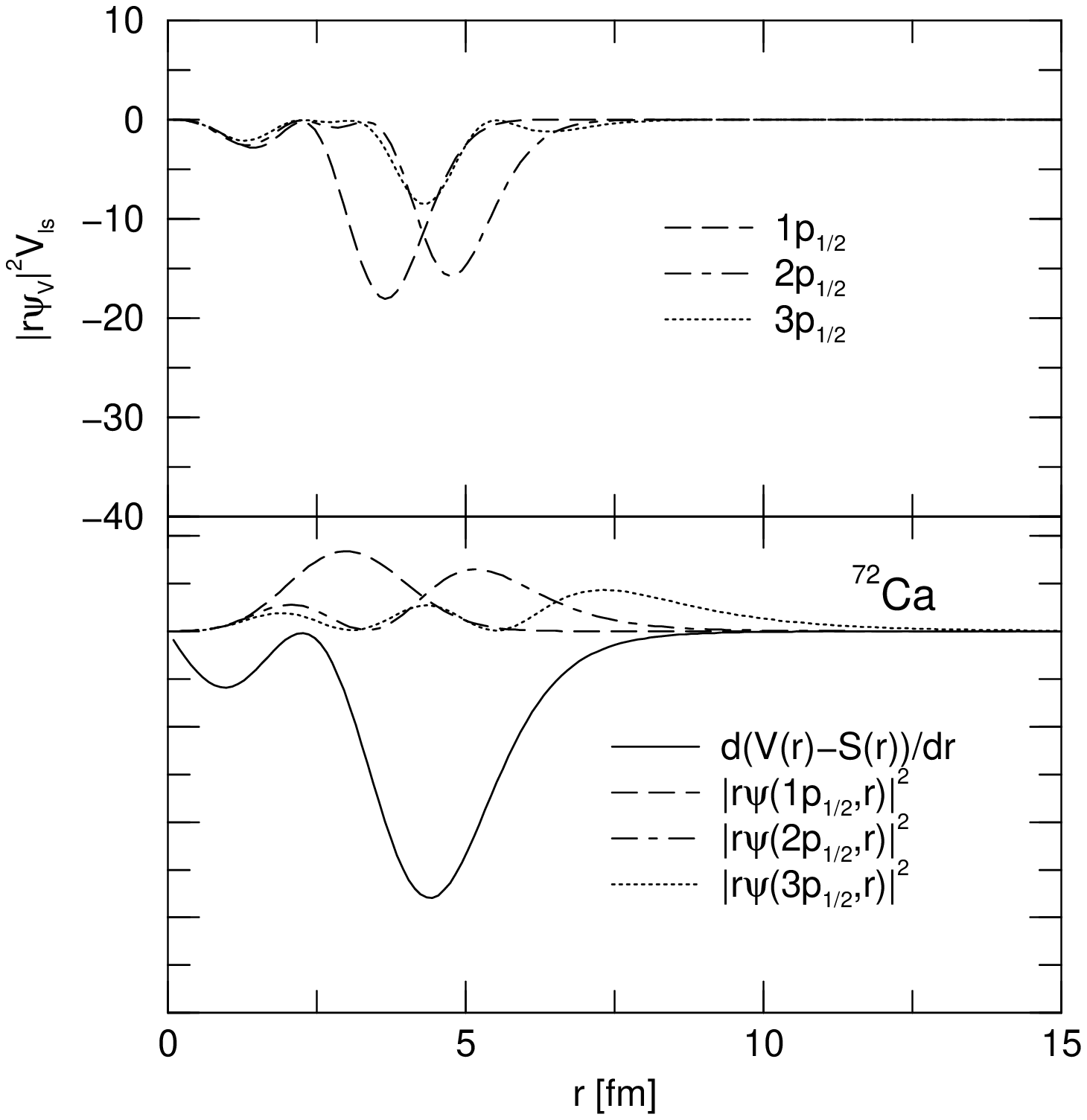}
\end{minipage}
\caption{ The overlap between the spin-orbit potential and density
distributions for $^{42}$Ca, $^{52}$Ca, $^{62}$Ca and $^{72}$Ca,
respectively. In each nucleus, the spin-orbit potential $V_{ls}$
multiplied by the density distribution of the $p_{1/2}$ orbit is
given in the upper panel; the derivative of anti-nucleon potential
and the density distributions for $p_{1/2}$ orbit in arbitrary
unit is given in the lower panel. } \label{soslapn}
\end{figure}

Seen from the right panel of Fig. \ref{sosQn}, the spin-orbit
splitting for $n=1$ states has similar features for $^{42}$Ca,
$^{52}$Ca, $^{62}$Ca and $^{72}$Ca. The splitting of the doublets
$(d_{1/2}, d_{3/2})$ is the largest one. With $l$, the splitting
for $n=1$ increases at first (from $p$ to $d$) and then decreases
(from $d$ to $f,~g$). This feature can also be understood with the
overlap of the spin-orbit potential and the density distribution
of these orbits. Here we just choose $^{72}$Ca as an example (Fig.
\ref{soslapl}), similar patterns also appear in other Ca isotopes.
The largest overlap happens for doublets $(d_{1/2}, d_{3/2})$ in
$^{72}$Ca is clearly seen from the upper panel of Fig.
\ref{soslapl}.

\begin{figure}[htbp]
\centering \includegraphics[width=0.5\textwidth]{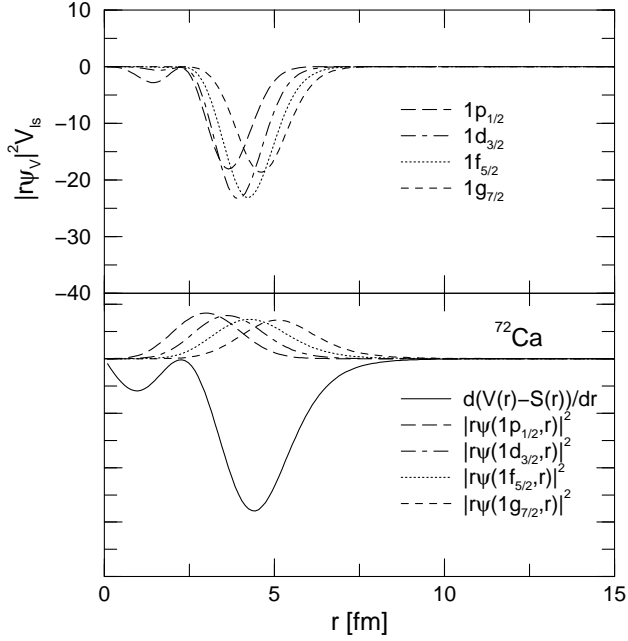}
\caption{The same as Fig. \ref{soslapn}, but for $n=1$ orbits in
$^{72}$Ca.} \label{soslapl}
\end{figure}

\section{The prospect of giant halos}

Nuclear halo phenomena have been studied and investigated by many
nuclear theorists and experimenters. More and more halo nuclei,
including neutron halo, proton halo, and halo at excited states,
have been identified and reported by different methods and
advanced instruments. However, only nuclei with one or two halo
nucleons are found by experiment until now.

Using various theoretical models, theorists can demonstrate
quantitatively the halo density distributions. Furthermore, the
giant halo nuclei with more halo nucleons have been predicted in
exotic Zr nuclei with $A>122$ using RCHB method \cite{MR98}. The
prediction of giant halo arouses a great interest of many nuclear
experimenters. Unfortunately, the exotic Zr nuclei are too heavy
to produce in nowadays accelerator. Here in this article, we
investigated the ground state properties of the whole Ca chain by
RCHB theory, and predict that giant neutron halo phenomena would
lie in exotic Ca nuclei with $A>60$ as well as the exotic Zr
nuclei. Compared to Zr isotopes, exotic Ca nuclei have lighter
mass number and would be much easier to synthesize. The heaviest
Ca isotope now known experimentally is $^{57}$Ca, so 5 more
neutrons are needed to form the  fringe giant halo nucleus
$^{62}$Ca, and 9 more neutrons to form the typical giant halo
nucleus $^{66}$Ca discussed above.

In order to confirm whether there exists a possibility of giant
halo nuclei in more wider mass region, in Fig. \ref{Fig.S2nNe-Al},
we show the two-neutron separation energies $S_{2n}$ for
even-neutron Ne, Na, Mg, and Al nuclei in drip line region. Open
symbols represent the values calculated from the RCHB theory with
the NL-SH parameter set while corresponding solid symbols
represent the data available. We note here that the calculations
were performed by keeping the spherical shape and hence the
comparison with the available experimental data are to be made
with care.  In Fig. \ref{Fig.S2nAr-Ti}, we plot $S_{2n}$ for
exotic even-neutron Ar, K, Ca, Sc and Ti isotopes. From these two
figures, the two-neutron separation energies $S_{2n}$ for all
these isotope chains are almost parallel. There are more than one
line lying within 2 MeV in the drip line region.  Therefore there
are quite a large mass region where a giant halo may exist.

\begin{figure}
\centerline{\epsfig{file=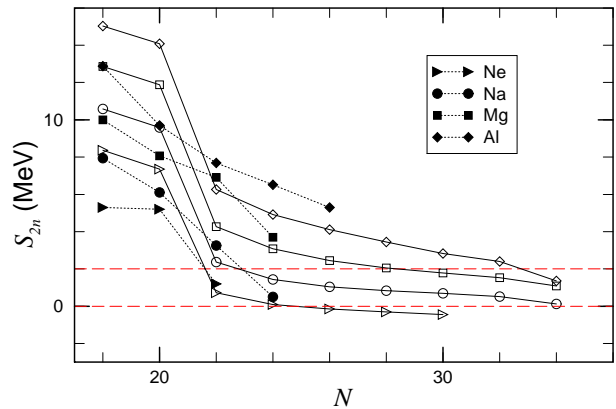,width=8.cm}} \caption{The
two-neutron separation energies $S_{2n}$ for even-neutron Ne, Na,
Mg, and Al nuclei in the drip line region. Open symbols represent
the values calculated with the RCHB theory with the NL-SH
parameter set, while corresponding solid symbols represent the
data available.  The horizontal line at 2MeV denotes the upper
limit for the possible occurrence of halos.} \label{Fig.S2nNe-Al}
\end{figure}

\begin{figure}
\centerline{\epsfig{file=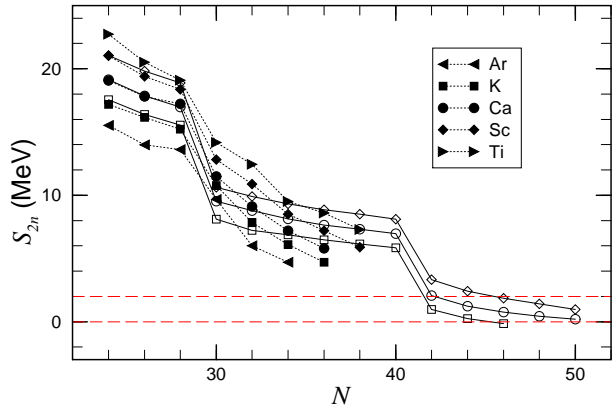,width=8.cm}} \caption{Same
as Fig.\ref{Fig.S2nNe-Al}, but for even Ar, K, Ca, Sc, and Ti
nuclei in the drip line region.}
 \label{Fig.S2nAr-Ti}
\end{figure}

Many nuclear physicists are making much effort to search the drip
line for heavier elements. A new experiment \cite{{No02},Lu02} has
been reported that both $^{37}$Na and $^{34}$Ne are bound, while
$^{33}$Ne and $^{36}$Na are missed in the experiment. Looking at
Fig. \ref{Fig.S2nNe-Al}, the $^{37}$Na and $^{34}$Ne nuclei lie in
the drip line region and approach to the giant halo nuclei.
Therefore we suggest that much more experimental effort should be
devoted to extend the study in this area and measure the mass to
find the possible giant halo nuclei. Similar situation can also be
seen in Fig.\ref{Fig.S2nAr-Ti} for Ar, K, Ca, Sc and Ti isotopes
near the neutron dripline.

\section{Summary}

We have investigated the ground state properties of even-even
proton magic O, Ca, Ni, Zr, Sn, and Pb isotopes with the
relativistic continuum Hartree-Bogoliubov (RCHB) theory.  We have
found good agreement with available experimental data for the
binding energies and the nuclear radii.  We have shown the binding
energies, $E_b$, two neutron separation energies, $S_{2n}$, and
root mean square radii. The predicted neutron drip nuclei are
$^{28}$O, $^{72}$Ca, $^{98}$Ni, $^{136}$Zr, $^{176}$Sn, and
$^{266}$Pb, respectively. Particularly the giant halos in the
neutron rich Ca and Zr isotopes close to the neutron drip line are
predicted. The giant halo properties in exotic Ca isotopes with
$A>60$ is discussed in detail.

Giant neutron halos in exotic Ca isotopes have been studied from
the analysis of $S_{2n}$, radii, nucleon density distribution,
single particle energy levels, occupation probabilities and the
contributions from the continuum. The spin-orbit splitting and the
potential diffuseness in Ca isotopes have also been investigated.
Summarizing the present investigation, we conclude:
\begin{enumerate}
\item Based on the analysis of two-neutron separation energies
$S_{2n}$, rms radii, single-particle levels spectra, the orbital
occupation, and the contribution of the continuum, giant halo
phenomena are suggested to appear in Ca isotopes with A$>$60.
Similar phenomena can also be seen for nuclei near the Na or Ar
isotopes near the neutron dripline. \item The neutron drip line
nucleus for Ca is $^{72}$Ca instead of $^{70}$Ca, which is caused
by the disappearance of the $N=50$ magic number due to the halo
effect of the $3s_{1/2}$ orbit. \item The giant halos developed in
these nuclei are due to the pairing correlation and the
contribution from the continuum e.g. the $3s_{1/2}$ orbit. \item
The spin-orbit splitting in Ca isotopes decreases monotonically
from the proton drip line to the neutron drip line for most cases.
This tendency mainly comes from the diffuseness of nuclear
potential with the neutron number.
\end{enumerate}

In this paper, the proton magic even-even nuclei from the proton
drip line to the neutron drip line are studied in detail in RCHB
theory. The important contribution from the continuum due to the
pairing correlations has been taken into account. The powerfulness
of the RCHB method has been demonstrated for the proton magic
nuclei using the assumption of the spherical shape. One of the
other important degrees of freedom for the exotic nuclei is the
deformation. The theoretical framework for exotic nuclei with the
deformation and contribution from continuum are in progress and
will be completed soon.

\section{Acknowledgements}
This work was partly supported by the Major State Basic Research
Development Program Under Contract Number G2000077407 and the
National Natural Science Foundation of China under Grant No.
10025522, 10047001, and 19935030. J.M. is grateful to RCNP of
Osaka University and Physikdepartment, Technische Universit\"at
M\"unchen for their support and warm hospitality, where a part of
this work was worked out.


\end{document}